%% file: main.tex
\documentclass[sigplan,screen]{acmart} 
\input{misc/copyright}

\input{misc/config}

\input{misc/macro}

\input{misc/authors}
\renewcommand{\shortauthors}{Jinjun Yi et al.}

\begin{document}



\title[\sysname: Accelerating LLM Decoding via Prefix-Aware Attention]{\sysname: Accelerating LLM Decoding via \underline{P}refix-\underline{A}ware A\underline{t}tention with Resource Efficient Multi-Tile Kernel}

\input{content/0_abstract}
\maketitle 

\input{content/1_introduction}

\input{content/2_background}

\input{content/3_motivation}
\input{content/4_1_overview}

\input{content/4_2_packing}
\input{content/4_3_forward}

\input{content/4_4_merge}
\input{content/5_1_eval_setup}
\input{content/5_2_eval_kernel}

\input{content/5_3_eval_e2e}
\input{content/5_4_eval_ablation}

\input{content/5_5_eval_overhead}
\input{content/6_related_conclusion}

\input{content/_appendix}

\bibliographystyle{ACM-Reference-Format}
\balance
\bibliography{misc/ref.bib}

\end{document}

%% file: misc/copyright.tex
\copyrightyear{2026}
\acmYear{2026}
\setcopyright{cc}
\setcctype{by}
\acmConference[ASPLOS '26] {Proceedings of the 30th ACM International Conference on Architectural Support for Programming Languages and Operating Systems, Volume 2}{March 21--26, 2026}{Pittsburgh, PA, USA.}
\acmBooktitle{Proceedings of the 30th ACM International Conference on Architectural Support for Programming Languages and Operating Systems, Volume 2 (ASPLOS '26), March 21--26, 2026, Pittsburgh, PA, USA}
\acmPrice{}
\acmISBN{979-8-4007-2359-9/2026/03}
\acmDOI{10.1145/3779212.3790200}

\ccsdesc[500]{Computing methodologies~Machine learning}
\ccsdesc[500]{Computer systems organization~Cloud computing}

\keywords{
Large Language Models;
LLM Inference;
Prefix Aware Attention;
GPU Kernel Scheduling
}

%% file: misc/config.tex
\usepackage{color}
\usepackage{hyperref} 
\usepackage{bm}
\usepackage{xspace}
\usepackage{pifont}
\usepackage{makecell}
\usepackage{tikz}   
\usepackage{multirow}
\usepackage{subcaption}
\usepackage{enumitem}
\usepackage{microtype}
\usepackage[ruled,vlined]{algorithm2e}

\usepackage[T1]{fontenc}
\usepackage[utf8]{inputenc}
\usepackage{listings}
\usepackage{xcolor}
\hypersetup{
    colorlinks=true,
    linkcolor=blue,
    filecolor=magenta,      
    urlcolor=cyan,
}

\makeatletter
\g@addto@macro{\UrlBreaks}{\UrlOrds}
\makeatother

\makeatletter
\def\@fnsymbol#1{%
  \ifcase#1\or
    \ensuremath{\dagger}\or 
    \ensuremath{\ast}\or
    \ensuremath{\ddagger}\or
    \ensuremath{\mathsection}\or
    \ensuremath{\mathparagraph}\or
    \ensuremath{\|}\else
    \@ctrerr
  \fi}
\makeatother

%% file: misc/macro.tex
\newcommand{\sysname}{{PAT}\xspace}
\newcommand{\sysnamewithnospace}{{PAT}}

\providecommand\parab[1]{\noindent\textbf{#1}}
\providecommand\parae[1]{\textbf{\textit{#1}}}

\newcommand{\ie}{\emph{i.e.,}\xspace}
\newcommand{\eg}{\emph{e.g.,}\xspace}

\newcommand{\secref}[1]{\S\ref{#1}}
\newcommand{\figref}[1]{Figure~\ref{#1}}
\newcommand{\tabref}[1]{Table~\ref{#1}}
\newcommand{\algoref}[1]{Algorithm~\ref{#1}}
\newcommand{\eqnref}[1]{Equation~\ref{#1}}

\newcommand{\appendixref}[1]{Appendix~\ref{#1}}

\newcommand{\num}[1]{%
  \tikz[baseline=(c.base)]%
    \node[circle, draw=violet, line width=0.5pt,
          inner sep=0pt, minimum size=8pt,
          font=\footnotesize, text=violet] (c) {#1};%
}

%% file: misc/authors.tex
\author{Jinjun Yi}
\orcid{0009-0002-5318-1434}
\authornote{Both authors contributed equally to this work.}
\affiliation{%
  \institution{Tianjin University}
  \city{Tianjin}
  \country{China}
}
\email{march_h@tju.edu.cn}

\author{Zhixin Zhao}
\orcid{0009-0009-7865-4905}
\authornotemark[1]
\affiliation{%
  \institution{Tianjin University}
  \city{Tianjin}
  \country{China}
}
\email{zhao612@tju.edu.cn}

\author{Yitao Hu}
\orcid{0009-0004-0458-0900}
\authornote{Corresponding author.}
\affiliation{%
  \institution{Tianjin University}
  \city{Tianjin}
  \country{China}
}
\email{yitao@tju.edu.cn}

\author{Ke Yan}
\orcid{0009-0002-5444-9797}
\affiliation{%
  \institution{Tianjin University}
  \city{Tianjin}
  \country{China}
}
\email{yank@tju.edu.cn}

\author{Weiwei Sun}
\orcid{0009-0000-9054-4017}
\affiliation{%
  \institution{Tianjin University}
  \city{Tianjin}
  \country{China}
}
\email{sww@tju.edu.cn}

\author{Hao Wang}
\orcid{0000-0002-1444-2657}
\affiliation{%
  \institution{Stevens Institute of Technology}
  \city{Hoboken}
  \state{NJ}
  \country{USA}
}
\email{hwang9@stevens.edu}

\author{Laiping Zhao}
\orcid{0000-0003-1967-2192}
\affiliation{%
  \institution{Tianjin University}
  \city{Tianjin}
  \country{China}
}
\email{laiping@tju.edu.cn}

\author{Yuhao Zhang}
\orcid{0000-0002-0118-8135}
\affiliation{%
  \institution{Tianjin University}
  \city{Tianjin}
  \country{China}
}
\email{yuhaozhang@tju.edu.cn}

\author{Wenxin Li}
\orcid{0000-0001-8507-0339}
\affiliation{%
  \institution{Tianjin University}
  \city{Tianjin}
  \country{China}
}
\email{toliwenxin@tju.edu.cn}

\author{Keqiu Li}
\orcid{0000-0003-1758-3030}
\affiliation{%
  \institution{Tianjin University}
  \city{Tianjin}
  \country{China}
}
\email{keqiu@tju.edu.cn}

%% file: content/0_abstract.tex
\begin{abstract}

LLM serving is increasingly dominated by decode attention, which is a memory-bound operation due to massive KV cache loading from global memory. Meanwhile, real-world workloads exhibit substantial, hierarchical shared prefixes across requests (\eg system prompts, tools/templates, RAG). Existing attention implementations fail to fully exploit prefix sharing: \textit{one-query-per-CTA} execution repeatedly loads shared prefix KV cache, while \textit{one-size-fits-all} tiling leaves on-chip resources idle and exacerbates bubbles for uneven KV lengths. These choices amplify memory bandwidth pressure and stall memory-bound decode attention.

This paper introduces \sysname, a prefix-aware attention kernel implementation for LLM decoding that organizes execution with a pack-forward-merge paradigm. \sysname packs queries by shared prefix to reduce repeated memory accesses, runs a customized multi-tile kernel to achieve high resource efficiency. It further applies practical multi-stream forwarding and KV splitting to reduce resource bubbles. The final merge performs online softmax with negligible overhead. We implement \sysname as an off-the-shelf plugin for vLLM. Evaluation on both real-world and synthetic workloads shows that \sysname reduces attention latency by 53.5\% on average and TPOT by 17.0-93.1\% under the same configurations against state-of-the-art attention kernels. \sysname's source code is publicly available at \url{https://github.com/flashserve/PAT}.

\end{abstract}



%% file: content/1_introduction.tex
\section{Introduction}
\label{sec:introduction}


Transformer-based Large Language Models (LLMs) have rapidly advanced across diverse domains, including conversational assistants~\cite{dam2024complete, dong2023towards}, code generation~\cite{anthropic2025claudecode, hui2024qwencoder}, retrieval-augmented generation (RAG)~\cite{lewis2020retrieval, yao2025cacheblend}, and agent/tool workflows~\cite{microsoft2025autogen, yuan2024easytool}. As deployment grows, latency and resource efficiency of online inference have become increasingly critical. While prior work has explored optimizations in memory management~\cite{kwon2023pagedattention, xu2025ellm}, scheduling~\cite{zhong2024distserve, sun2024llumnix}, quantization~\cite{lin2024awq, chee2023quip}, and architecture~\cite{zhou2022mixture, liu2024deepseek}, there remain new challenges and opportunities. We highlight two emerging trends below.


\parae{Longer contexts and outputs.} Context length has scaled to millions of tokens~\cite{meta2025llama4}, while techniques like Chain-of-Thought (CoT)~\cite{wei2022chain} demand longer outputs. As a result, decode attention (\ie attention operations in the decoding stage) increasingly dominates end-to-end latency due to repeated loading of growing KV caches from global to on-chip memory, which poses a memory-bound challenge.


\parae{Shared prefixes across requests.} System prompts, RAG documents, and agent templates introduce multi-level shared prefixes across requests, creating an optimization opportunity. Existing systems~\cite{zheng2024sglang, kwon2023pagedattention} implement prefix KV cache reuse, which reduces memory footprint by storing and reusing KV cache across requests. However, it can \textit{not} reduce global memory accesses, which is the bottleneck for decode attention, thereby leading to higher attention latency.

Recent attention kernel implementations aim to reduce attention latency~\cite{shah2024flashattention3, ye2025flashinfer, ye2024chunkattention, pan2025fasttree, song2024tackling, zhu2024relayattention}, but still face redundant KV cache loads from global memory and inefficient resource utilization. \textit{Query-centric} kernels~\cite{dao2022flashattention, dao2023flashattention2, ye2025flashinfer, song2024tackling} follow a one-query-per-CTA strategy to map each query and corresponding KV cache into an independent Cooperative Thread Array (CTA; also known as a thread block) for execution on the GPU, causing redundant global memory loads for shared prefixes. \textit{KV-centric} kernels~\cite{pan2025fasttree, zhu2024relayattention, yao2024deft, ye2024chunkattention} instead pack the KV cache of cross-query shared prefix into one CTA to reduce redundant memory accesses, but adopt a one-size-fits-all design that fixes the tile size of the GPU kernel~\cite{nvidia2025cudabestpractices} and applies padding to fill the tile, which wastes on-chip memory and limits CTA concurrency.

Based on the analysis, we argue that \textbf{an efficient attention kernel implementation for LLM decoding should be memory-oriented with prefix-aware execution}, so as to reduce redundant memory accesses and maintain high resource efficiency. However, the dynamicity of workloads makes this design non-trivial. First, the combination of multi-level shared prefixes and the dynamic join-and-leave nature of continuous batching makes the structure of shared KV caches in a decode batch (\ie batched queries in decode stage) variable across decode steps. The attention kernel must pack the decode batch into CTAs effectively, so as to reduce redundant global memory accesses with low runtime overhead. Second, the autoregressive nature of LLMs makes queries have diverse KV lengths, and the number of queries that share the same KV blocks in one CTA varies over time. These features bring dynamic hardware resource requirements per CTA, which in turn affect the memory bandwidth usage and GPU utilization. The attention kernel must adapt to the dynamicity to achieve high resource efficiency.

To address these challenges, we present \sysname, an attention kernel implementation for LLM decode stage with a \textit{pack–forward–merge} execution paradigm. Specifically, in the \textit{pack stage} (\secref{sec:design_pack}), \sysname employs a prefix-aware pack scheduler and a lazy update mechanism, which efficiently pack decode batches into CTAs to reduce redundant global memory accesses with negligible overhead. It further designs a multi-tile kernel based on resource-efficiency analysis, so as to adapt to CTA dynamicity and avoid on-chip memory waste. In the \textit{forward stage} (\secref{sec:design_forward}), \sysname designs a multi-stream forward and long-KV split strategy, which parallelizes the multi-tile kernel across multiple CUDA streams and splits CTAs with excessively long KV lengths. These designs enable \sysname to eliminate execution bubbles and achieve high global memory bandwidth utilization. In the \textit{merge stage} (\secref{sec:design_merge}), \sysname applies a lightweight kernel with online softmax~\cite{dao2022flashattention}, which merges partial results across CTAs for each query.

In summary, this paper makes the following contributions:
\begin{enumerate}[leftmargin=20pt,labelsep=6pt,topsep=3pt]
    \item We provide an in-depth analysis of the bottlenecks in decode attention (\secref{sec:background}) and the optimization opportunities and challenges introduced by shared prefixes (\secref{sec:motivation}), from both hardware characteristics and the attention execution pipeline perspectives.
    \item We design and implement a prefix-aware attention kernel \sysname for LLM decoding. It incorporates several novel designs within the pack–forward–merge paradigm to reduce redundant global memory accesses and achieve efficient hardware utilization (\secref{sec:design_overview}-\secref{sec:design_merge}).
    \item We evaluate \sysname on synthetic and real-world workloads, demonstrating that compared with state-of-the-art baselines, it reduces attention latency by 53.5\% on average under the same decode batches (\secref{sec:eval_kernel_perf}) and lowers TPOT by 17.0-93.1\% under the same request rate (\secref{sec:eval_e2e}).
\end{enumerate}


%% file: content/2_background.tex
\section{Background}
\label{sec:background}

\subsection{LLM Inference and Attention}
\label{sec:background_llm_inference}




Transformer-based Large Language Models (LLMs) follow an autoregressive inference process with two phases: prefill and decode. In the prefill phase, the model performs a full forward pass over the input sequence to generate the first token. The decode phase then iteratively generates one token at a time until reaching an end-of-sequence token or a preset length. Each decode step involves three major computations: Self-Attention, Query-Key-Value-Output (QKVO) projection, and Multi-Layer Perceptron (MLP). Among these, Self-Attention is central for modeling global dependencies as:
$$
    \text{Attention}(Q, K, V) = \text{softmax}\left(\frac{
        {\color{orange}\boldsymbol{QK^T}}
    }{\sqrt{d_k}}\right)
    {\color{orange}\boldsymbol{V}}
$$


During decoding, the query $Q$ from the current token attends to all previous keys $K$ and values $V$. Since $K$ and $V$ do not change across steps, they are stored in global memory using a Key-Value (KV) Cache, reducing redundant computation but increasing memory footprint.

\begin{figure}[t]
    \centering
    \subfloat[end-to-end latency]{
        \includegraphics[width=0.23\textwidth]{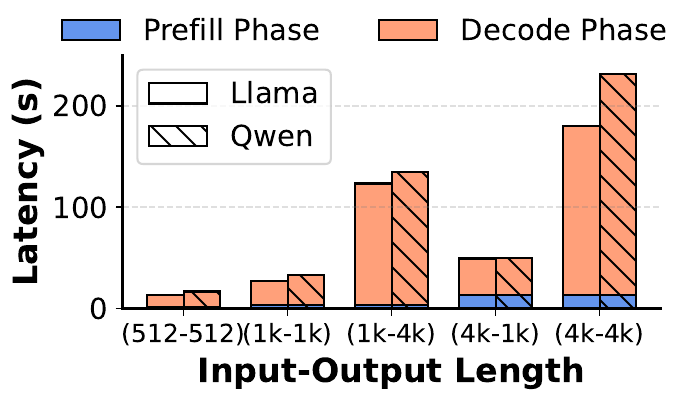}
        \label{fig:background_latency_PD}
    } 
    \subfloat[decode phase latency]{
        \includegraphics[width=0.23\textwidth]{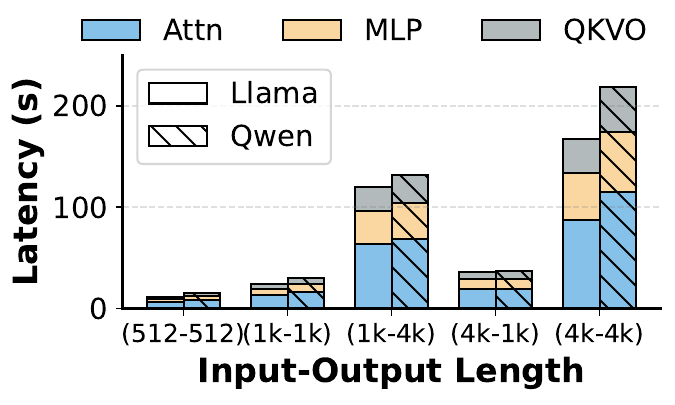}
        \label{fig:background_latency_decode}
    }
    \caption[]{Latency breakdown of Llama-3-8B and Qwen3-8B across context length on A100, vLLM v0.9.0, batch size 64}
    \label{fig:background_latency_breakdown}
    \Description{}
\end{figure}

\subsection{Memory Bottleneck of Attention}
\label{sec:background_attn_bottleneck}


Driven by the inference scaling law~\cite{wu2025inference}, LLMs are trending toward \textit{longer contexts} and \textit{longer outputs}. For instance, Llama-4 supports up to 10 million input tokens~\cite{meta2025llama4}, and Chain-of-Thought prompting~\cite{wei2022chain} has significantly increased output lengths for complex reasoning.


Although KV Caching avoids repeated computation, it has shifted the bottleneck from compute to memory access: each decode step must fetch a growing amount of KV cache from global memory to on-chip memory~\cite{zadouri2025hardware, behnam2025rocketkv}. As input and output lengths grow, this access cost dominates inference latency. As in \figref{fig:background_latency_breakdown}, decode attention can contribute up to 53\% of the total latency for Llama-3-8B and Qwen3-8B.


\subsection{GPU Execution Model}
\label{sec:background_attn_execution}

\begin{figure}
    \centering
    \includegraphics[width=0.42\textwidth]{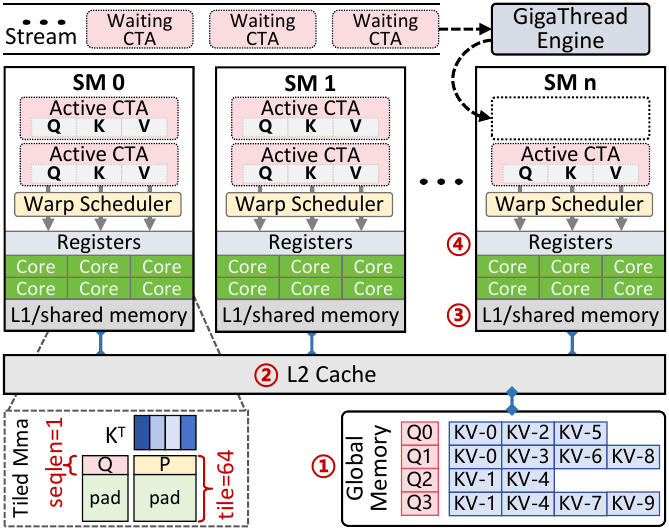}
    \caption{GPU architecture of typical NVIDIA GPUs.}
    \label{fig:background_gpu_arch}
    \Description{}
\end{figure}

To fully understand the decode attention bottleneck, we should examine the GPU's hardware and software architecture, as well as the execution process of the attention kernel. As shown in \figref{fig:background_gpu_arch}, the GPU hardware architecture is composed of \num{1} \textbf{global memory}, \num{2} \textbf{L2 Cache}, an array of Streaming Multiprocessors (SMs), and a global hardware scheduler (GigaThread Engine). Each SM is a basic computational unit containing various CUDA Cores, Tensor Cores, and its own hierarchical memory structure: a programmer-managed \num{3} \textbf{Shared Memory / L1 Cache} that enables low-latency data exchange among threads within a Cooperative Thread Array (CTA, aka Thread Block), and a \num{4} \textbf{Register File} for thread-private storage. As shown in \tabref{tab:background_gpu_memory}, the GPU memory hierarchy involves a significant trade-off between size and speed, with global memory access being orders of magnitude slower than on-chip memory access.


\begin{table}[t!]
    \centering
    \footnotesize
    \setlength{\tabcolsep}{1pt}
    \begin{tabular*}{\columnwidth}{@{\extracolsep{\fill}} c ccccc}
        \toprule
        \textbf{Level}                          & \textbf{Shared By}    & \textbf{Size} & \textbf{Latency} & \textbf{Bandwidth$^*$} & \textbf{Type} \\
        \midrule
        Register                                & Thread                & 256KB/SM$^{\dag}$      & $\sim$2ns        & $\sim$20 TB/s      & on-chip \\
        \makecell{Shared Memory \\/ L1 Cache}   & CTA                   & 192KB/SM$^{\ddag}$      & $\sim$20ns       & $\sim$19 TB/s      & on-chip \\
        L2 Cache                                & All SMs               & 40MB           & $\sim$140ns      & $\sim$2 TB/s       & on-chip \\
        Global Memory                           & All SMs               & 80GB        & $\sim$200ns      & $\sim$2 TB/s       & off-chip \\
        \bottomrule
        \multicolumn{6}{l}{$^*$Read/write bandwidth from the upper memory level.} \\
        \multicolumn{6}{l}{$^{\dag}$Each thread is limited to 255 registers (4*8 bits each).} \\
        \multicolumn{6}{l}{$^{\ddag}$Each CTA can address up to 163 KB of shared memory.} \\
    \end{tabular*}
    \caption{Memory hierarchy of A100-SXM4-80GB~\cite{nvidia2020a100, abdelkhalik2022demystifying}.}
    \label{tab:background_gpu_memory}
\end{table}


\begin{figure}
    \centering
    \includegraphics[width=0.49\textwidth]{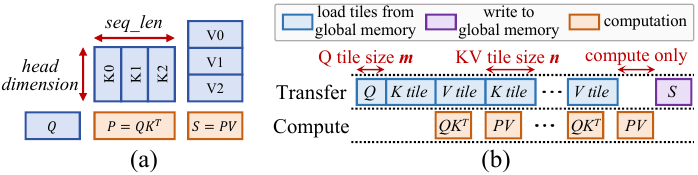}
    \caption{(a) Two General Matrix-Vector multiplications (GEMV) in attention. (b) Tiling execution pipeline.}
    \label{fig:background_attention_pipeline}
    \Description{}
\end{figure}


When executing an attention kernel, the workload is divided into CTAs, scheduled across SMs. During this process, the required KV cache data is transferred from \textit{slow global memory} up the hierarchy into \textit{fast on-chip shared memory}, where it is finally loaded into registers for computation. State-of-the-art implementations~\cite{dao2022flashattention, dao2023flashattention2, ye2025flashinfer} are designed to exploit this memory hierarchy. For example, FlashAttention~\cite{dao2022flashattention, nvidia2025cudabestpractices} partitions K/V caches into small tiles along the sequence length (\figref{fig:background_attention_pipeline}a) and processes them in a pipelined fashion: while one tile is computed, the next is prefetched asynchronously (\figref{fig:background_attention_pipeline}b), reducing memory latency.


Nonetheless, decode attention remains bottlenecked by two challenges: limited bandwidth between global and on-chip memory, and low arithmetic intensity due to heavy KV cache loading. Thus, further optimization must follow two principles: (1) \textit{Reduce KV cache transfer loads from global memory}, and (2) \textit{Fully utilize available memory bandwidth}.

%% file: content/3_motivation.tex
\section{Motivation}
\label{sec:motivation}

In this section, we first characterize the shared-prefix pattern commonly observed in LLM workloads (\secref{sec:motivation_prefix}) and identify two limitations in existing attention implementations under this pattern: redundant global memory accesses (\secref{sec:motivation_memory}) and low hardware utilization (\secref{sec:motivation_bubble}). We then introduce the pack–forward–merge paradigm to address these problems (\secref{sec:motivation_insight}) and analyze the corresponding challenges (\secref{sec:motivation_challenge}).

\subsection{Shared Prefixes and Prefix Reuse}
\label{sec:motivation_prefix}

\begin{figure}[t]
    \centering
    \includegraphics[width=0.47\textwidth]{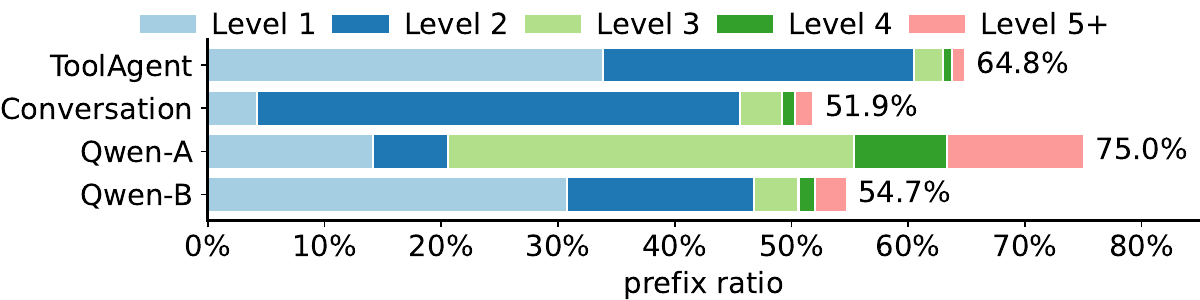}
    \caption{Prefix ratio of four traces: ToolAgent: tool and agent interaction workload~\cite{qin2025mooncake}; Conversation: online conversation workload~\cite{qin2025mooncake}; Qwen-A: online API service~\cite{wang2025kvcache}; Qwen-B: task automation with API calling workload~\cite{wang2025kvcache}.}
    \label{fig:motivation_trace_profile}
    \Description{}
\end{figure}


Shared prefixes between requests are common and hierarchical in modern LLM workloads~\cite{openai2025promptcaching, wang2020generalizing, jin2024dvd}. As shown in \figref{fig:motivation_trace_profile}, our analysis of four real-world traces~\cite{qin2025mooncake, wang2025kvcache} finds a prefix ratio of $51.9-75.0\%$, meaning that more than half of KV cache tokens come from prefixes reused across requests. Under continuous batching, this cross-request sharing leads to \textit{multi-level intra-batch prefixes}. On the Conversation and ToolAgent traces, using the setup in \secref{sec:eval_setup}, intra-batch shared prefixes cover $2.8-82.6\%$ of KV caches, and each batch ends up with 2.72 distinct shared prefixes on average. These significant and hierarchical shared prefixes present unique opportunities and challenges for LLM inference optimization.

Systems like SGLang\cite{zheng2024sglang} and vLLM\cite{kwon2023pagedattention} adopt \textit{prefix reuse} by mapping shared logical prefixes to a single physical copy in global memory. This approach reuses KV caches across requests to reduce memory usage. However, it can \textit{not} leverage intra-batch shared prefixes to reduce global memory accesses, which remain the bottleneck. To understand why, we analyze two key inefficiencies in existing approaches under shared-prefix scenarios: redundant memory access (\secref{sec:motivation_memory}) and underutilized hardware resources (\secref{sec:motivation_bubble}).

\subsection{Redundant Memory Access}
\label{sec:motivation_memory}

\begin{figure}[t]
    \centering
    \includegraphics[width=0.48\textwidth]{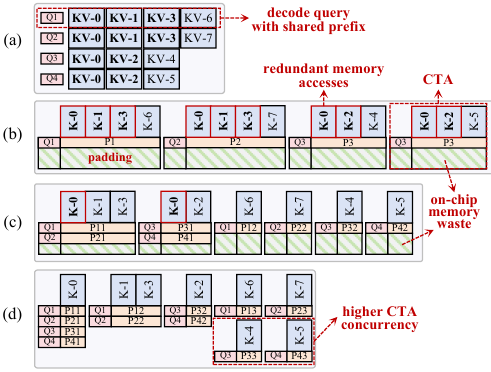}
    \caption{Comparison of packing strategies. (a) A decode batch of 4 queries with shared prefixes. (b) Query-centric packing causes redundant memory access. (c) KV-centric packing causes memory waste. (d) Memory-centric prefix-aware packing avoids redundancy and improves utilization.}
    \label{fig:motivation_packing}
    \Description{}
\end{figure}

A major inefficiency in current attention kernels stems from redundant memory accesses. To execute a decode batch, attention kernels use a packing strategy that groups queries and their KV caches into CTAs. In the decode batch with 4 queries (\figref{fig:motivation_packing}a), existing kernels' query-centric paradigm adopts a \textit{one-query-per-CTA} packing strategy~\cite{shah2024flashattention3, ye2025flashinfer}, where each query and its KV are independently assigned to a CTA (\figref{fig:motivation_packing}b). While simple to schedule, this strategy causes shared KV prefixes (\eg KV-0, KV-1) to be repeatedly loaded from slow global memory into on-chip memory. Although L2 cache offers partial reuse, it is limited by size and bandwidth.

\begin{figure}[t]
    \centering
    \subfloat[redundant memory accesses]{
        \includegraphics[width=0.22\textwidth]{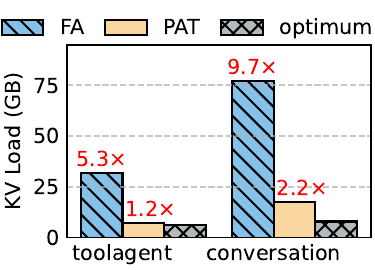}
        \label{fig:motivation_latency_memory}
    } 
    \subfloat[resource inefficiencies]{
        \includegraphics[width=0.25\textwidth]{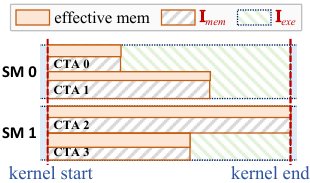}
        \label{fig:motivation_bubble_pipeline}
    } 
    \caption[]{Limitations of existing attention kernels. (a) Average KV cache load from global memory per decode step on the \textit{toolagent} and \textit{conversation} traces, comparing FlashAttention (FA), \sysname, and the theoretical optimum. (b) Two-dimensional resource inefficiencies.} 
    \label{fig:motivation_limitation}
    \Description{}
\end{figure}




To examine the inefficiency of the one-query-per-CTA strategy, we profile FlashAttention~\cite{shah2024flashattention3} using \texttt{ncu}~\cite{nvidia2025nsightcompute}, comparing its KV cache traffic with the theoretical minimum (where each shared block is loaded once), and with \sysname. As shown in \figref{fig:motivation_latency_memory}, FlashAttention incurs 4.3-8.7$\times$ more KV cache than the theoretical minimum, and 4.1–7.5$\times$ more than \sysname. This result confirms significant overhead from redundant memory access for the one-query-per-CTA paradigm.


\parae{Observation \#1}: Existing query-centric attention kernels suffer from substantial memory access redundancy due to their one-query-per-CTA execution paradigm.

\subsection{Two-Dimension Resource Inefficiencies}
\label{sec:motivation_bubble}


Furthermore, existing attention kernels exhibit inefficient hardware utilization. Both query-centric and KV-centric designs use a tiled pipeline (\figref{fig:background_attention_pipeline}) to overlap memory access and computation. In this paradigm, selecting appropriate tile sizes (\ie query tile size $m$ and KV tile size $n$ from \figref{fig:background_attention_pipeline}b) is critical for resource efficiency~\cite{dao2022flashattention}. However, existing kernels adopt a \textit{one-size-fits-all} design, which specifies a single, hard-coded tile size for all CTAs (\eg $m=64, n=32$~\cite{shah2024flashattention3}).



As shown in \figref{fig:motivation_packing}b-c, this static approach ignores the dynamic nature of LLM workloads, leading to significant resource inefficiencies along two dimensions\footnote{While specific prior works~\cite{pan2025fasttree} design one more query tile sizes $m$, it is still insufficient for handling highly dynamic query numbers and KV lengths.} in \figref{fig:motivation_bubble_pipeline}: (1) Memory Waste ($\mathbb{I}_{mem}$): When fewer than $m$ queries share a KV prefix, CTAs must pad inputs, wasting shared memory and registers to store unused data. (2) Execution Bubble ($\mathbb{I}_{exe}$): Due to varying KV lengths across CTAs, fixed-size tiling causes imbalanced workloads, leaving SMs underutilized in the tail stages of execution.



\parae{Observation \#2}: Existing query-centric and kv-centric attention kernels suffer from two-dimension resource inefficiencies due to their one-size-fits-all execution paradigm.


\subsection{Insight: Pack-forward-merge Paradigm}
\label{sec:motivation_insight}


Building on the observations above, we identify two design principles for efficient decode attention (\figref{fig:motivation_packing}d): (1) Intra-CTA KV Cache Sharing: pack queries with shared KV prefixes into the same CTA to enable KV reuse in shared memory and avoid redundant global memory access. (2) Resource-Efficient Kernel Design: tailor kernel implementations to GPU architecture and CTA configurations to sustain high memory bandwidth and minimize resource inefficiencies.


Following these principles, we propose \textit{pack–forward–merge paradigm} as follows: (1) \textbf{Pack}: group queries by shared prefix into CTAs to eliminate redundant KV loads; (2) \textbf{Forward}: execute CTAs using resource-optimized kernels that output partial results in tiles; (3) \textbf{Merge}: apply online softmax to combine partial results into final outputs.


\subsection{Challenges}
\label{sec:motivation_challenge}

Although the pack–forward–merge paradigm directly targets the objectives of reducing global memory access and improving resource utilization, it faces two major challenges:


\textbf{Challenge 1: Packing complexity.} LLM inference often involves deep, multi-level shared prefixes and long contexts, significantly expanding the packing search space. Each prefix level may yield multiple packing candidates with different trade-offs. Additionally, continuous batching introduces dynamic request changes, requiring frequent packing updates. An effective strategy must account for both prefix hierarchy and batch dynamism, generating CTA assignments with minimal latency (\secref{sec:design_pack.pack_scheduler}).


\textbf{Challenge 2: Workload variability.} Autoregressive decoding leads to large variation in KV lengths. Grouping queries by shared prefixes further amplifies variation in CTA sizes, ranging from one to dozens of queries. This variability affects resource demands and execution time across CTAs. The forward stage must use a kernel design that adapts to both hardware and workload characteristics (\secref{sec:design_pack.multi_tile}) and employs scheduling strategies to minimize time bubbles (\secref{sec:design_forward}).

%% file: content/4_1_overview.tex
\section{Overview}
\label{sec:design_overview}

To address these challenges, we design \sysname, a \textit{memory-centric attention kernel implementation} that follows the pack-forward-merge paradigm. It serves as a backend for the serving system vLLM~\cite{kwon2023pagedattention}. In the \textit{pack} stage, \sysname adopts a profit-model-based heuristic packing strategy to aggregate queries that share KV into a CTA, so as to mitigate redundant global memory accesses (\secref{sec:design_pack.pack_scheduler}). It further designs multi-tile kernels and a runtime tile selector to choose an efficient kernel for each CTA (\secref{sec:design_pack.multi_tile}). In the \textit{forward} stage, \sysname adopts multi-stream execution and a long-KV split strategy to enable efficient kernel execution, which reduces execution bubbles (\secref{sec:design_forward}). Finally, in the \textit{merge} stage, \sysname uses a lightweight kernel based on online softmax to merge each query’s intermediate results across CTAs and produce the final output (\secref{sec:design_merge}).

%% file: content/4_2_packing.tex
\section{Pack Scheduler}
\label{sec:design_pack}

\begin{figure*}[t]
    \centering
    \includegraphics[width=0.975\textwidth]{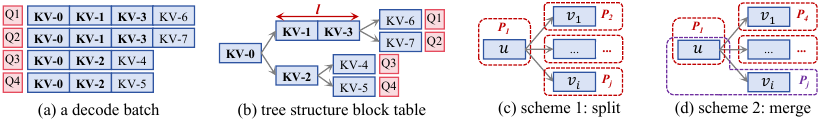}
    \caption{Workflow of the pack scheduler: (a) An input decode batch with 4 queries. (b) Tree structure block table. (c) Packing scheme 1 that splits leaf nodes with the parent node. (d) Packing scheme 2 that merges specific leaf node with the parent node.}
    \label{fig:design_pack.block_table}
    \Description{}
\end{figure*}

We first introduce a pack scheduler that packs a decode batch into CTAs by shared prefixes to reduce redundant global memory accesses (\secref{sec:design_pack.pack_scheduler}), and then present the customized multi-tile kernels that efficiently execute these CTAs (\secref{sec:design_pack.multi_tile}).


\subsection{Pack Scheduler}
\label{sec:design_pack.pack_scheduler}

\parab{Insight and Approach.} As noted in \secref{sec:motivation_memory}, the one-query-per-CTA paradigm repeatedly loads KV blocks for shared prefixes, worsening the memory bottleneck of decode attention. We therefore introduce a heuristic pack scheduler that (i) abstracts a decode batch’s block table into a prefix tree, (ii) scores candidate packing schemes with a memory-centric profit model, and (iii) packs the decode batch into memory-optimized CTAs to cut redundant global accesses.

\parab{Problem Formulation} Since decode attention is memory-bound (\secref{sec:background_attn_execution} and \secref{sec:motivation_memory}), the pack stage aims to minimize global memory accesses for a given decode batch (\figref{fig:design_pack.block_table}a). The \textit{input} is a batch of queries plus a block table where each row lists block IDs for a query; a shared prefix appears as identical leading block IDs across rows. The \textit{output} is a partition $\mathcal{P}=\{P_1,P_2,\dots\}$ of CTAs, where each $P_i$ packs queries sharing successive identical prefix blocks (queries may be split across CTAs). We seek $\mathcal{P}^\ast$ that minimizes memory accesses, comprised of loaded KV cache size and per-query intermediate reads/writes due to splits/merges. The search space grows exponentially with query count and prefix lengths~~\cite{pan2025fasttree}, so an exact solver is impractical for online serving, and we turn to a heuristic pack scheduler described next.

\parab{Tree Structure Block Table.}
For efficient implementation, the pack scheduler first converts the two-dimensional block table (\figref{fig:design_pack.block_table}a) into a tree structure block table (\figref{fig:design_pack.block_table}b). Each internal node represents a shared prefix of KV blocks. It has two attributes: (1) $l$, the KV-cache length of this shared prefix, and (2) $s$, the number of queries that share it. Each leaf corresponds to one query, and the path from the root to that leaf reconstructs the query’s full KV cache blocks.

\parab{Intra-node profit.} We first discuss the profit and overhead of packing the queries within a single non-leaf node $u$ (with attributes $l_u$ and $s_u>1$) into one CTA. First, compared with the one-query-per-CTA paradigm, packing $s_u$ queries with shared KV length of $l_u$ into a single CTA could reduce the KV cache loads from $s_u$ times to $1$ time. Therefore, it saves $(s-1)\,l_u\,d$ global memory accesses ($d$ is the head dimension). However, packing the shared prefixes of all queries into a CTA produces $2\,s_u$ times (half comes from node $u$ and the other half from leaf nodes) per-query intermediate writes and reads. Therefore the overhead of memory access\footnote{To ensure numerical accuracy, intermediate results are stored in FP32, so the overhead is multiplied by 2.} is $2\times (2\,s_u\,d + 2\,s_u\,d)\,=\,8\,s_u\,d$. Then, the profit–overhead ratio for packing a non-leaf node to a CTA is:
$$
r=\frac{(s_u-1)\,l_u\, d}{8\,s_u\,d}\ \ge\ \frac{l_u}{16}.
$$
In practice, the length of shared KV $l_u \geq 16$ since sharing is performed at the granularity of KV blocks~\cite{kwon2023pagedattention, shah2024flashattention3}, whose sizes are typically larger than 16. Therefore, \textit{packing a node into a CTA yields a positive profit.}

\parab{Inter-node profit.} When child nodes are involved, the profit and overhead change. Let $\{v_1, v_2, ... v_i\}$ be the children of $u$, where child $v_i$ has KV length $l_i$ and queries $s_i$ ($s_u=\sum_i s_i$). We compare two schemes in \figref{fig:design_pack.block_table}c-d.

\parae{Scheme 1: Split.} As in \figref{fig:design_pack.block_table}c, a naive packing scheme is splitting each node into an individual CTA. Following the analysis above, we can derive the overall profit as:
\begin{equation}
\label{eq:scheme1}
\underbrace{(s_u-1)\,l_u\, d}_{\text{profit of $u$}}
\,-\,\underbrace{4 s_u d}_{\text{overhead of $u$}}
\,+\,\underbrace{\textstyle \sum_i (s_i-1)\,l_i\, d}_{\text{profit of $\{v_1, v_2, ... v_i\}$}}.
\end{equation}

\parae{Scheme 2: Merge.} As shown in \figref{fig:design_pack.block_table}d, when considering child nodes, we can pack specific child node $v_i$ and parent node $u$ into a CTA to eliminate their intermediate results. In this case, the number of queries associated with node $u$ becomes $s_u' = s_u - s_i$, while the KV length $l_u$ remains unchanged. Then the profit of the unmerged part could also be estimated using \eqnref{eq:scheme1}, and the profit of the merged part $u\!\sim\! v_i$ becomes $(s_i-1)\,(l_u+l_i)\, d$. Therefore, the overall profit is:
\begin{align}
\label{eq:scheme2}
& \underbrace{(s_u-s_i-1)\,l_u\, d}_{\text{profit of $u$}}
\;-\;\underbrace{4 (s_u-s_i) d}_{\text{overhead of $u$}} \notag\\
&\quad +\ \underbrace{\textstyle\sum_{k\neq i} (s_k-1)\,l_k\, d}_{\text{profit of $\{v_1, v_2, ... v_{i-1}\}$}}
\ +\ \underbrace{(s_i-1)\,(l_u+l_i)\, d}_{\text{profit of $u\!\sim\!v_i$}}.
\end{align}

\parae{Scheme comparison.} The incremental profit of Scheme~2 over Scheme~1 is $4 s_i d \;-\; l_u d$. Hence Scheme~2 is preferred when $4s_j>l_u$. When the shared prefix at $u$ is short and the specific child node $v_i$'s queries are large enough, merging them achieves higher profit by eliminating unnecessary intermediate results. Otherwise, keep $u$ and $v_i$ separate.




\parab{Pack Scheduler.} We implement a heuristic scheduler guided by the profit analysis to pack the decode batch into CTAs. It first converts the block table into a forest, where each root is a unique first-level prefix and each root–leaf path encodes a query’s multi-level shared prefixes plus its non-shared suffix. Each node stores the shared block IDs and the query IDs. For each tree, the scheduler invokes \textit{TreeHeuristic} (\algoref{alg:design_pack.tree_heuristic}) to produce CTAs and adds them to $\mathcal{P}$. \textit{TreeHeuristic} packs each leaf as an independent CTA (line 4), scans the children of each internal node, applies the inter-node profit model to choose a scheme, and recursively packs children (lines 5–13). It then packs the node’s remaining queries into a CTA and returns with its children’s CTAs (lines 14–15). This yields memory-efficient CTAs with fewer global-memory accesses and linear complexity $O(|V|{+}|E|)$ since each node and edge is processed once\footnote{$V$ and $E$ denote numbers of nodes and edges in tree-structured block table.}.

\parab{Lazy Update.} Although the pack scheduler is linear, its overhead can grow with batch size and number of KV blocks. We mitigate this with a lazy-update strategy that (1) reuses a scheduling result across continuous-batching iterations until the block table changes (\eg request arrivals/departures or new KV block assignments) and (2) moves the scheduler into the serving system and runs it asynchronously once the block table is available. These reduce scheduler invocations from once per transformer layer to once several continuous-batching iterations and overlap scheduling with pre-attention task (\eg metadata preparation, QKV projection), thereby substantially reducing the exposed scheduling latency (see \secref{sec:eval_overhead}) without affecting model accuracy.



\subsection{Multi-tile Kernel}
\label{sec:design_pack.multi_tile}

\begin{figure*}[t]
    \centering
    \subfloat[memory latency]{
        \includegraphics[width=0.235\textwidth]{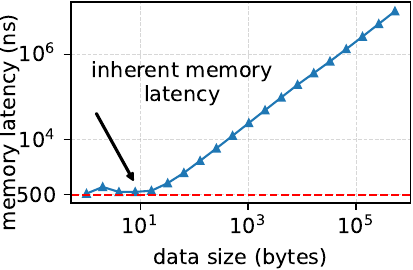}
        \label{fig:design_pack.memory_latency}
    } 
    \subfloat[tile size configurations]{
        \includegraphics[width=0.235\textwidth]{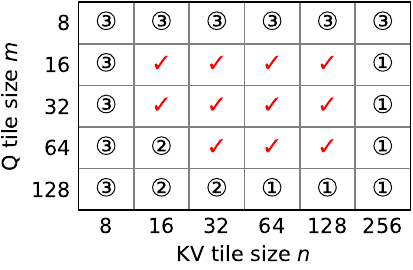}
        \label{fig:design_pack.kernel_table}
    } 
    \subfloat[bandwidth equivalence]{
        \includegraphics[width=0.235\textwidth]{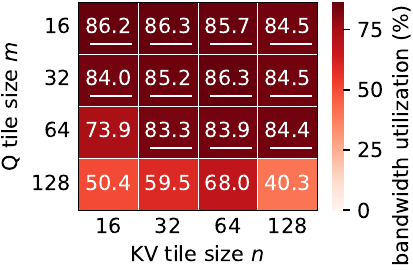}
        \label{fig:design_pack.kernel_eq_bw}
    } 
    \subfloat[latency equivalence]{
        \includegraphics[width=0.235\textwidth]{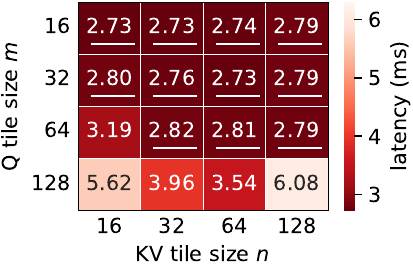}
        \label{fig:design_pack.kernel_eq_lat}
    }
    \caption[]{Multi-tile kernel design and validation on A100-SXM4-80GB GPU. (a) Global-to-shared memory transfer latency for varying data sizes (averaged over 1,000 reads). (b) Offline-selected tile-size configurations; check marks denote feasible settings, and circled numbers indicate violated constraints. (c) Average bandwidth utilization under different tile sizes. (d) Average kernel latency under different tile sizes.} 
    \label{fig:design_pack.kernel}
    \Description{}
\end{figure*}


\parab{Insight and Approach.} Given packed CTAs, \sysname must choose per-CTA Q-tile $m$ and KV-tile $n$ to maximize resource efficiency. Because query sizes and KV lengths vary widely, single–tile-size kernels with padding (\eg prior works~\cite{dao2023flashattention2, ye2025flashinfer, pan2025fasttree, zhu2024relayattention}) cannot adapt and suffer poor utilization (\secref{sec:motivation_bubble}). To better exploit GPU shared memory and registers, we introduce a \textit{customized tiled attention kernel design} that consists of: (1) a multi-tile kernel suite, where feasible $(m,n)$ configurations are derived from offline hardware and CTA-constraint analysis and implemented as resource-efficient kernels; and (2) a tile-size selector, an online decision-tree that selects the per-CTA $(m,n)$ to balance performance and parallelism.

\input{content/_algorithm}




\parab{Multi-tile kernel.} Tile sizes (Q tile $m$ and KV tile $n$) critically affect Tensor Core efficiency: they determine a CTA’s shared-memory and register demand, which in turn constrain resident CTA concurrency and active warps. To obtain resource-efficient kernels, we derive three key constraints that significantly reduce the $(m,n)$ search space.

\parae{\ding{172} Register and shared-memory constraints} (upper bounds on $m,n$). Increasing $m$ or $n$ raises CTA shared-memory and register usage and can induce register spilling~\cite{nvidia2025cudaprogguide}. To keep the kernel within hardware limits, we enforce two bounds: (1) \textit{Shared-memory constraint.} One CTA’s shared-memory usage comprises the Q tile, K/V tile, and intermediate results (data type size $b'$, usually higher precision). It must not exceed per-SM shared memory $S_{\mathrm{smem}}$:
$$m\,h\,b \;+\; n\,h\,b \;+\; m\,h\,b' \;\le\; S_{\mathrm{smem}}.$$
(2) \textit{Register constraint.} We bound per-thread\footnote{Commonly 255 32-bit registers per thread on recent GPUs.} and aggregate register use to avoid spilling: per-thread registers $\le S_{\mathrm{reg\_thr}}$; total registers of concurrent CTAs on an SM $\le S_{\mathrm{register}}$. Because compiler effects make analytic estimates unreliable, we obtain per-thread $R_{\mathrm{thr}}(m,n)$ and per-CTA $R_{\mathrm{CTA}}(m,n)$ via offline compilation and static analysis, and enforce:
$$
R_{\mathrm{thr}}(m,n) \le S_{\mathrm{reg\_thr}},\qquad
C \cdot R_{\mathrm{CTA}}(m,n) \le S_{\mathrm{register}}.
$$

\parae{\ding{173} High bandwidth utilization} (lower bound for $n$). To saturate global memory bandwidth, the pipeline must \textit{keep enough data in flight} to cover the inherent memory latency. \figref{fig:design_pack.memory_latency} shows transfer latency vs. data size: the flat region gives the inherent latency $L$ (ns) and the linear region the sustainable bandwidth $B$ (Bytes/ns). Thus the in-flight data $D_{\text{flight}}$ must satisfy $D_{\text{flight}}\ge L\! \times\! B$. In the attention pipeline $D_{\text{flight}}$ is the total size of all K or V tiles being loaded by concurrently resident CTAs, \ie $D_{\text{flight}}=S\,C\,n\,h\,b$ where $S$ is the number of SMs, $C$ the concurrent CTAs per SM, $h$ the head dimension, and $b$ the KV datatype size (bytes). Rearranging gives the lower bound, which complements the CUTLASS-derived constraints and ensures the memory bus remains well utilized:
$$
n \;\ge\; \left\lceil \frac{L\,B}{S\,C\,h\,b}\right\rceil,
$$



\parae{\ding{174} CUTLASS constraint} (lower bounds for $m$ and $n$): Efficient use of CUTLASS/CuTe MMA requires both tile sizes to be powers of two and at least\footnote{It depends on data format; e.g., the minimum tile size is 32 for int8.} 16~\cite{cutlass_quickstart}:
$$
m, n \in \{2^k \mid k\in\mathbb{N},\, 2^n \ge 16\}
$$



\parae{Put it together.} Based on \ding{172} to \ding{174}, an offline configuration solver is designed to compute feasible tile size ($m$, $n$) pairs per hardware target, thereby providing kernels that execute efficiently for dynamic CTAs. \figref{fig:design_pack.kernel_table} shows a set of available tile size configuration under A100-80GB.



\parab{Tile Selector.} Given the packed CTAs and feasible multi-tile kernel configurations, the tile selector assigns a tile size configuration pair $(m, n)$ to each CTA at runtime using a set of rules. These rules are derived offline based on analysis of (1) when different tiles behave equivalently, and (2) how $m$ and $n$ affect resource efficiency.




\begin{figure}[t]
    \centering
    \subfloat[bandwidth equivalence]{
        \includegraphics[width=0.235\textwidth]{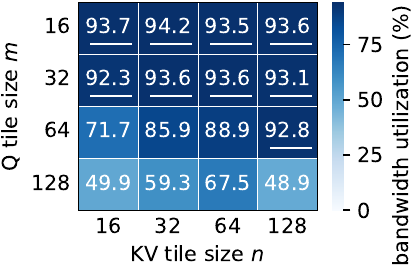}
        \label{fig:design_pack.kernel_eq_bw_h100}
    } 
    \subfloat[latency equivalence]{
        \includegraphics[width=0.235\textwidth]{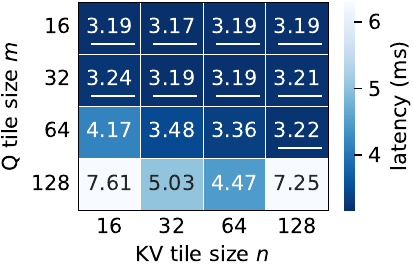}
        \label{fig:design_pack.kernel_eq_lat_h100}
    }
    \caption[]{Multi-tile kernel validation on H100-SXM4-80GB GPU. (a) Average bandwidth utilization under different tile sizes. (b) Average kernel latency under different tile sizes.}
    \label{fig:design_pack.kernel_h100}
    \Description{}
\end{figure}

\parae{Kernel equivalence.}
Per the offline constraint solving above, the candidate $(m,n)$ configurations sustain high bandwidth utilization, and decode-attention latency is dominated by global memory bandwidth. Thus, for decode batches \emph{without} shared prefixes or execution bubbles, these configurations are \emph{performance-equivalent} in bandwidth and latency. To validate this, we executed a decode batch without prefixes (KV length $1024$) under various configurations and used batch size $1134$, which is a common multiple of CTA concurrency across configurations on A100, avoiding execution bubbles. As shown in \figref{fig:design_pack.kernel_eq_bw} and \figref{fig:design_pack.kernel_eq_lat}, all candidate configurations (underlined) sustain $83\%$-$86\%$ bandwidth utilization and exhibit similar end-to-end latency (difference $<2\%$). This demonstrates that, in the absence of prefixes and bubbles, varying the tile configuration within the feasible set does not change CTA performance.

Porting \sysname's multi-tile kernel to other GPUs only requires re-deriving equivalent tile size configurations using constraints~\ding{172}-\ding{174}. On Hopper H100-SXM-80GB GPUs, this procedure removes the $(64,32)$ and $(64,64)$ configurations from \figref{fig:design_pack.kernel_table}, and the remaining entries form the equivalent kernel set. \figref{fig:design_pack.kernel_h100} reports validation on H100 at batch size $1188$, a common multiple of CTA concurrency across all configurations\footnote{Batch size $1134$ for A100 and $1188$ for H100 are only used in the kernel-equivalence validation, not in the evaluation.}. All equivalent configurations achieve $92.3\%$-$94.2\%$ bandwidth utilization and similar kernel latency. These results indicate that the same constraint-based procedure generalizes across architectures.



\parae{Deriving Q tile $m$.} The shared prefixes make the query size per CTA dynamic. To mitigate padding-induced memory waste $\mathbb{I}_{mem}$ on the $m$ dimension, the selector uses a \textit{round-up} rule: given CTA’s query size $q$, choose the smallest $m$ in feasible q tile sizes with $m \ge q$. For instance, when $q=20$, it chooses $m=32$ rather than $16$, since $m=16$ will split the query into two CTAs and result in redundant accesses of the shared KV cache. Larger and performance-equivalent tile sizes such as $64$ or $128$ are also avoided to preserve on-chip memory for KV tile size $n$ selection as follows.

\begin{figure}[t]
    \centering
    \includegraphics[width=0.47\textwidth]{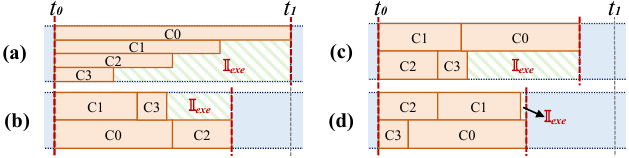}
    \caption{Execution pipeline of four CTAs under different concurrency. (a) High concurrency with dynamic KV lengths yields low tail efficiency and a large execution bubble $\mathbb{I}_{exe}$. (b–d) Lower concurrency cuts per-CTA latency and thus reduces $\mathbb{I}_{exe}$ across the shown execution orders.}
    \label{fig:design_pack.bubble_pipeline}
    \Description{}
\end{figure}






\parae{Deriving KV tile $n$.} The choice of $n$ must adapt to KV length. \textit{For long KV}, high CTA concurrency causes severe execution bubbles and poor tail efficiency (\figref{fig:design_pack.bubble_pipeline}a), so we prefer larger $n$: this raises per-CTA on-chip memory, reduces per-SM concurrency $C$, thereby increasing bandwidth available per CTA and shrinking the execution bubble $\mathbb{I}_{exe}$ across different execution orders (\figref{fig:design_pack.bubble_pipeline}b-d). \textit{For short KV}, the last tiling iterations can make the compute-only portion significant (right side of \figref{fig:background_attention_pipeline}b), so a smaller $n$ shortens the final tile and avoids a compute bubble is preferred. \eg at KV length 192, $n{=}128$ yields $(192-128)/128\approx50\%$ compute bubble in the last tile, while $n{=}64$ removes it and is faster due to kernel equivalence. Guided by these trade-offs, we profile each candidate $n$ offline by sweeping KV length to derive the largest performance-equivalent tile sizes until the choice stabilizes; the resulting mapping is encoded as a piecewise decision tree. During online serving, tile selector performs a constant-time lookup per CTA to choose the profiled $n$.



%% file: content/_algorithm.tex
{\footnotesize

\LinesNumbered    
\SetNlSty{}{}{:}  
\SetAlgoNlRelativeSize{0.5}  


\begin{algorithm}[t]
\caption{TreeHeuristic}
\label{alg:design_pack.tree_heuristic}
\KwIn{Root node $root$, corresponding KV $blocks$}
\KwOut{Packs $\mathcal{P}$ from the tree}
$\mathcal{P} \leftarrow \emptyset$\;
\If{$\mathrm{IsLeaf}(r)$}{
    \textcolor{blue}{// Pack the non-shared KV into a CTA}\\
    \Return{$\mathrm{Pack}(root.\mathrm{query}, blocks)$}\;
}
\For{$c \in \mathrm{Children}(root)$}{
    \textcolor{blue}{// Use profit model to choose the scheme}\\
    \If{$4\times c.size < root.len_s$}{
        \textcolor{blue}{// Scheme 1: split root and child into separate CTAs}\\
        $\mathcal{P} \leftarrow \mathcal{P} \cup \mathrm{TreeHeuristic}(c, c.\mathrm{blocks})$\;
    }
    \Else{
        \textcolor{blue}{// Scheme 2: merge root's blocks with $c$'s blocks}\\
        $\mathcal{P} \leftarrow \mathcal{P} \cup \mathrm{TreeHeuristic}(c, c.\mathrm{blocks} \cup blocks)$\;
        $root.\mathrm{RemoveQuery}(c.\mathrm{queries})$\;
    }
}
\textcolor{blue}{// Pack remaining queries and KV blocks into a CTA}\\
$\mathcal{P} \leftarrow \mathcal{P} \cup \mathrm{Pack}(root.\mathrm{queries}, blocks)$\;

\Return{$\mathcal{P}$}\;
\end{algorithm}

}


%% file: content/4_3_forward.tex
\section{Multi Kernel Forward}
\label{sec:design_forward}



After \secref{sec:design_pack} packs the batch into CTAs and selects multi-tile kernels, we design two forward-stage strategies to mitigate kernel execution bubbles $\mathbb{I}_{\mathrm{exe}}$ (\secref{sec:motivation_bubble}).

\parab{Multi-Stream Forward.} Kernels with different tile size configurations launch and execute sequentially on the GPU, since the GPU requires static kernel launch parameters derived from tile size. 
This serial execution incurs resource inefficiencies due to frequent kernel launches and execution bubbles, with the latter accumulating across consecutive kernels. To address this, we create a separate CUDA stream for each distinct tile size configuration $(m,n)$ obtained from \secref{sec:design_pack.multi_tile}. The scheduler groups CTAs by their $(m,n)$ pair and enqueues each group into its corresponding stream, so that all CTAs with the same configuration execute sequentially within that stream, while different streams run in parallel. 
This design overlaps the launch overhead of subsequent kernels with the execution of preceding kernels and mitigates execution bubbles by kernel parallelism (\secref{sec:eval_ablation}).

\parab{Long KV Split.} Multi-stream execution alone cannot eliminate execution bubbles in all cases, because some CTAs may have KV lengths that are orders of magnitude larger than others. We therefore adopt a KV-dimension splitting strategy similar to~\cite{pan2025fasttree}. Specifically, we split any CTAs with KV length exceeding the mean KV length for all CTAs of the batch into equal parts to keep the KV length under the mean value. This shortens the completion time of the last finishing CTAs and improves overall SM utilization of kernel execution.


%% file: content/4_4_merge.tex
\section{Output Merge}
\label{sec:design_merge}


We implement a lightweight merge kernel that uses online softmax~\cite{dao2022flashattention} to combine partial results at per-query granularity. Each CTA produces three per-query and per-head intermediates: a max score, a log-sum-exp accumulator, and a partial value-weighted sum. The merge kernel loads these intermediates from global memory, reduces them with online softmax, normalizes the accumulated value-weighted sum, concatenates all heads, and writes the final query output back to global memory. The small global read/write overhead for these intermediates is accounted for in the pack scheduler’s overhead analysis when deriving the end-to-end packing scheme (\ref{sec:design_pack.pack_scheduler}).



%% file: content/5_1_eval_setup.tex
\section{Evaluation}
\label{sec:eval}

\begin{figure*}[t]
    \centering
    \includegraphics[width=\linewidth]{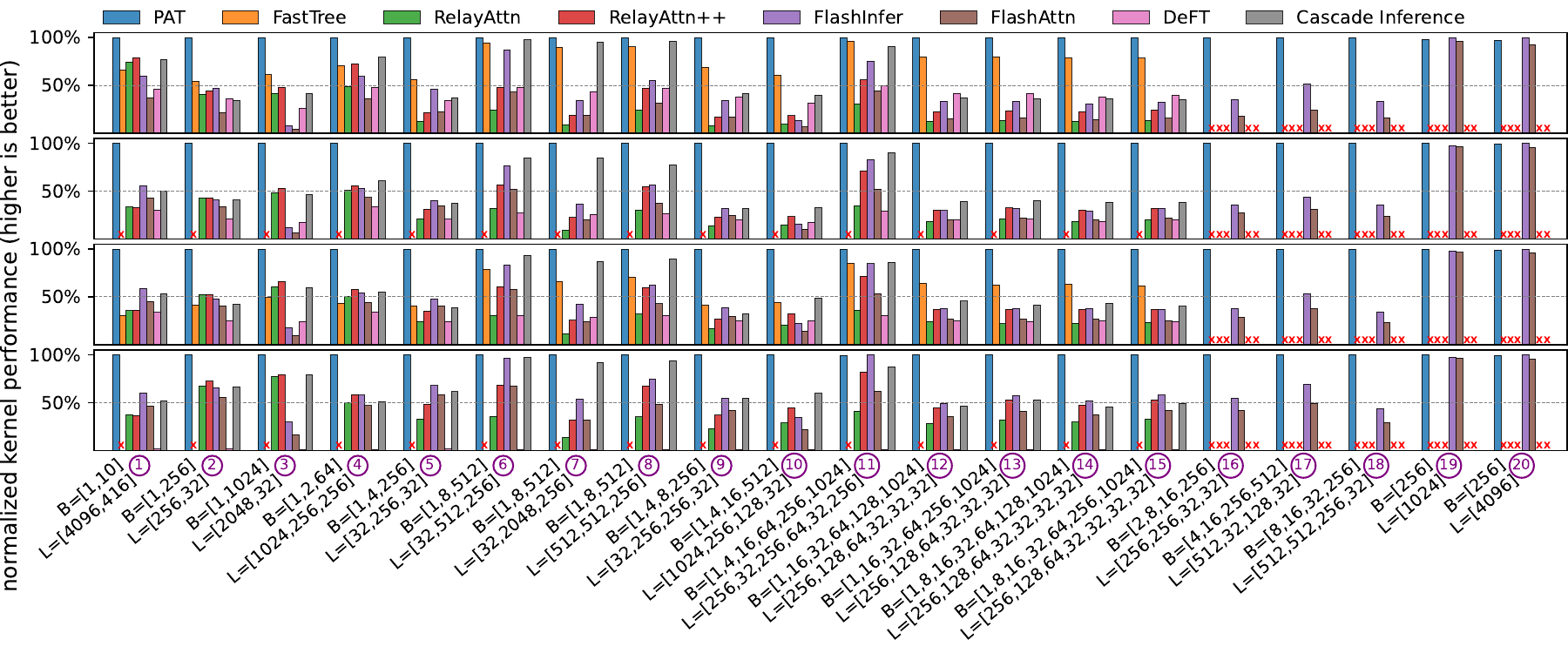}
    \caption{Normalized kernel performance (higher is better) of \sysname and the baselines for the attention computation across various decode batch configurations on NVIDIA A100 GPU (80GB). The four panels from top to bottom show head configurations ($num\_attention\_heads/num\_key\_value\_heads$) of 32/32, 16/8, 32/8, and 64/8. Missing bars arise because RelayAttention lacks support for multi-level or multiple first-level prefixes, and FastTree does not support the 16/8 and 64/8 head settings.}
    \label{fig:eval_kernel_overall}
    \Description{}
\end{figure*}

In this section, we present the implementation of \sysname (\secref{sec:eval_implementation}) and the experimental setup (\secref{sec:eval_setup}). We then conduct a set of experiments to answer three key questions:
\begin{enumerate}[leftmargin=20pt,labelsep=6pt,topsep=3pt]
    \item How does \sysname’s efficiency scale across diverse batch sizes, models, and prefix structures? (\secref{sec:eval_kernel_perf})
    \item What performance improvement does \sysname achieve for online serving under real-world workloads? (\secref{sec:eval_e2e}, \secref{sec:eval_dist_moe})
    \item What is the contribution of each feature of \sysname to the overall performance gain? (\secref{sec:eval_ablation})
    \item What is the impact of \sysnamewithnospace’s overhead? (\secref{sec:eval_overhead})
\end{enumerate}

\subsection{Implementation of \sysname}
\label{sec:eval_implementation}

We implement \sysname as a full attention kernel for the decode stage with about 3k lines of Cutlass/CuTe~\cite{nvidia2025cutlassdocs, nvidia2025cutequickstart} and C++ code. The multi-tile kernel (\secref{sec:design_pack.multi_tile}), multi-stream forward (\secref{sec:design_forward}), and merge kernel (\secref{sec:design_merge}) are built with Cutlass/CuTe. The asynchronous pack scheduler (\secref{sec:design_pack.pack_scheduler}) and API wrappers are implemented in C++. To overlap the data transfers with computation, all data movement from the global memory to the shared memory uses the \texttt{cp\_async} primitive, together with double buffering~\cite{nvidia2025cudaprogguide}. We expose the kernel API to Python via pybind11, and integrate it into vLLM~\cite{kwon2023pagedattention} (v0.9.0) as an off-the-shelf plugin with about 1.2k lines of Python code. \sysname treats vLLM's paged KV cache as its substrate: KV entries are managed as fixed-size blocks in block tables, and the pack scheduler operates only on block IDs produced by vLLM. This design lets \sysname reuse vLLM's existing KV paging implementation. To enable \sysname in vLLM, only an environment variable \texttt{VLLM\_ATTENTION\_BACKEND=\MakeUppercase{\sysname}} is required.


\subsection{Experiment Setup}
\label{sec:eval_setup}

\parab{Models and Testbed.}
For kernel benchmark (\secref{sec:eval_kernel_perf}), we evaluate \sysname on both NVIDIA A100 GPU (80GB) and NVIDIA H100 GPU. For end-to-end online serving (\secref{sec:eval_e2e}), we use two representative LLMs, Qwen3-8B and Llama-3-8B, on a single A100 GPU. We further evaluate \sysname under distributed settings and Mixture-of-Experts (MoE) architectures in \secref{sec:eval_dist_moe} using Qwen2.5-72B-Instruct on four A100 GPUs and Qwen3-30B-A3B on one A100 GPU. The software environment is CUDA 12.4 and PyTorch 2.7.0.


\parab{Baselines.} We compare against seven attention implementations spanning query-centric and KV-centric designs:
\begin{enumerate}[leftmargin=20pt,labelsep=6pt,topsep=3pt]
    \item \texttt{FlashAttention}~\cite{dao2022flashattention, dao2023flashattention2} (v2.5.9): query-centric; maps each query to a CTA with a fixed tile size config $(64,128)$.
    \item \texttt{FlashInfer}~\cite{ye2025flashinfer} (v0.2.5): query-centric; improves SM load balance by dynamic CTA partitioning; decoding tile config $(16,128)$.
    \item \texttt{FastTree}~\cite{pan2025fasttree}: KV-centric; uses a compute-oriented cost model to pack and reduce repeated KV loads; two tile configs $(64,32)$ and $(16,32)$.
    \item \texttt{RelayAttention}~\cite{zhu2024relayattention}: KV-centric; packs first-level shared prefixes into CTAs to cut redundant memory accesses and runs them with FlashAttention’s kernel.
    \item \texttt{RelayAttention++}: our extension of RelayAttention to exploit vLLM-style KV-cache reuse; it stores shared KV blocks from non-first-level prefixes in the same physical space so redundant KV loads can benefit from L2 cache, further improving performance (\secref{sec:eval_kernel_perf}).
    \item \texttt{DeFT}~\cite{yao2024deft}: KV-centric, aggregates queries with shared KV and adjusts the KV length in each CTA for load balance; fixed tile size config $(32, 16)$.
    \item \texttt{Cascade Inference}~\cite{zihao2024cascade}: KV-centric, packs prefixes into CTAs using fixed settings.
\end{enumerate}

\parab{Kernel-Performance Workloads (\secref{sec:eval_kernel_perf}).} To compare kernel performance under identical batch and prefix structures, we construct \textit{synthetic decode batches} as input, following~\cite{pan2025fasttree}. Each decode batch is specified by $B$ and $L$. $B$ defines the prefix-tree structure and the number of leaves (\ie batch size). \eg $B=[1,4,16]$ yields two shared-prefix levels with 1 and 4 nodes, and 16 leaves. $L$ gives KV lengths per level. \eg $L=[128,256,1024]$ sets level-1 and level-2 shared prefixes to 128 and 256 tokens, with 1024 non-shared tokens per request. We vary $(B, L)$ combinations to reflect different shared-prefix structures and batch settings as in \figref{fig:eval_kernel_overall}. Besides, we choose \textit{four head configurations} (\#heads, \#kv\_heads) common in Llama~\cite{meta2024llama3}, Qwen~\cite{yang2025qwen3}, and Gemma~\cite{google2025gemma} models: $(64,8)$, $(32,8)$, $(16,8)$, and $(32,32)$. The head dimension and data type are set to the commonly used 128 and FP16.


\parab{End-to-end workloads (\secref{sec:eval_e2e} and \secref{sec:eval_ablation}).} We evaluate \sysname under online serving using vLLM~\cite{kwon2023pagedattention} (v0.9.0) on two real-world traces. (1) \texttt{toolagent}~\cite{qin2025mooncake}: tool/agent workloads with task-specific system prompts (overall cache hit rate 59\%). (2) \texttt{conversation}: combines the Meta-AI system instruction (total lengths 2517/2522 tokens for Llama3/Qwen3 tokenizers) with burstgpt prompts, following prior work~\cite{pan2025fasttree}. We randomize language and country fields in the system instruction to create a three-level prefix whose lengths are 46, 348, 2123 (Llama3) or 45, 351, 2126 (Qwen3). We use only the first 30 minutes of each trace due to cost limit.

\begin{figure*}[t]
    \centering
    \includegraphics[width=\linewidth]{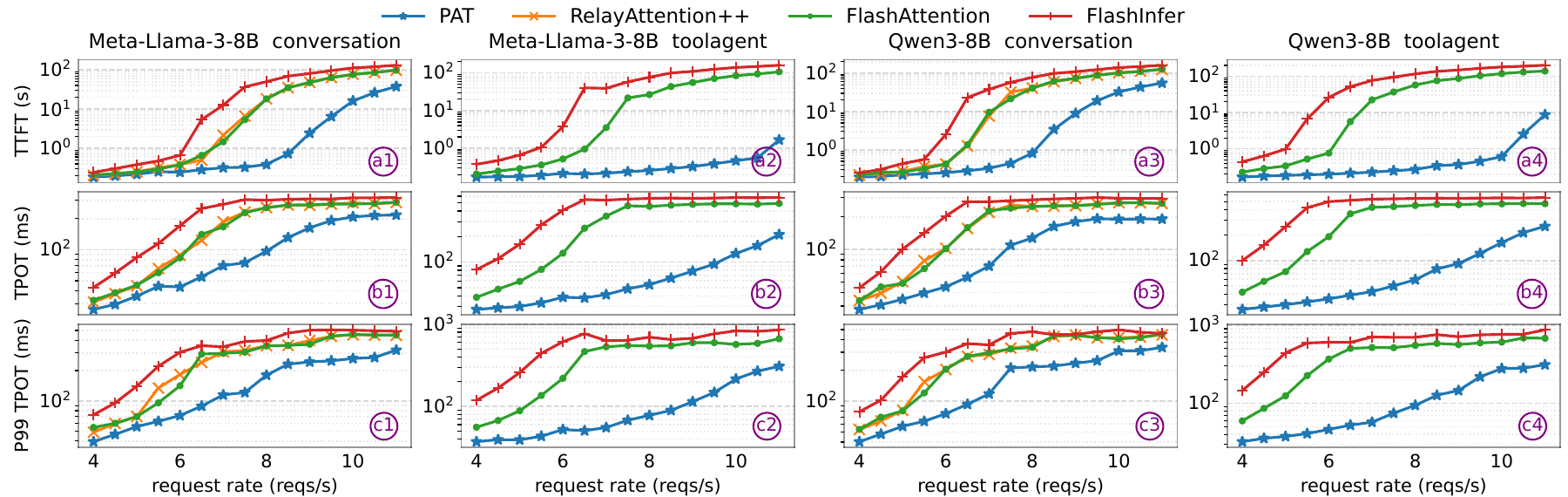}
    \caption{End-to-end performance of \sysname and the baselines under two models and two traces. Note that RelayAttention++ lacks support for multiple first-level prefixes, so its results on \texttt{toolagent} trace are unavailable.}
    \label{fig:eval_e2e_overall}
    \Description{}
\end{figure*}

\parab{Metrics.} For kernel performance, we primarily compare attention latency under varied input configurations. For each configuration, we run 20 repetitions and report the average completion latency. For end-to-end comparison, we focus on three metrics: average request completion latency, time to the first token (TTFT), and time per output token (TPOT). 


%% file: content/5_2_eval_kernel.tex
\subsection{Kernel Performance}
\label{sec:eval_kernel_perf}

\parab{Overall results.} \figref{fig:eval_kernel_overall} reports normalized kernel performance (metric $\text{latency}^{-1}$, normalized to \sysname) across decode-batch configurations on NVIDIA A100 GPU (80GB)\footnote{Evaluation results on NVIDIA H100 GPU are presented in \appendixref{sec:appendix_kernel_perf}.}. For configurations with shared prefixes (\num{1}–\num{18}), \sysname achieves up to $21.5\times$, $11.7\times$, $3.2\times$, $11.9\times$ and $5.7\times$ speedups over FlashAttention, FlashInfer, FastTree, RelayAttention and RelayAttention++, respectively, across four attention-head settings. These gains arise from three factors: (1) a prefix-aware packing scheduler that cuts redundant global KV accesses (\secref{sec:design_pack.pack_scheduler}); (2) a multi-tile kernel combined with an online tile-size selector that adapts $(m,n)$ per CTA to better use bandwidth and on-chip memory (\secref{sec:design_pack.multi_tile}); and (3) multi-stream forward execution and long-KV splitting that mitigate execution bubbles caused by multi-tile kernels and KV-length dynamics (\secref{sec:design_forward}). Together, these designs make \sysname consistently more efficient than the baselines.


\parab{Compared with query-centric kernels.} Against query-centric FlashAttention and FlashInfer, \sysname reduces attention latency by $67.8\%$ and $52.1\%$ on average under configurations with prefixes (\num{1}–\num{18}). The gap widens with larger batch size or longer shared prefixes (\eg \num{4}–\num{5} and \num{6}–\num{7}) because the one-query-per-CTA design of query-centric kernels forces repeated global KV loads that grow more costly in those cases. We also include configurations \num{19} and \num{20} in \figref{fig:eval_kernel_overall}, which remove shared prefixes and thus represent workloads without prefix reuse. In this case, \sysname no longer reduces global memory accesses but still benefits from its multi-tile kernel and multi-stream forward. As a result, \sysname achieves $1.6\%$ lower latency on average with fewer execution bubbles.


\parab{Compared with KV-centric kernels.} FastTree, which also targets multi-level shared prefixes, remains the strongest baseline but is still $3.8\%$–$68.9\%$ slower than \sysname for two reasons: (1) its compute-oriented packing cost model is ill-suited to memory-bound decode attention (\secref{sec:background_attn_execution} and \secref{sec:motivation_memory}); and (2) its double-tile approach launches two kernels serially as in \figref{fig:ablation_CTA_pipeline_serial}, introducing execution bubbles. By contrast, \sysname uses a memory-oriented packing strategy plus multi-stream forward to avoid these inefficiencies. RelayAttention++ cuts latency by $67.4\%$ versus RelayAttention, confirming that L2 plus KV-cache reuse reduces redundant global loads, yet RelayAttention++ is still $1.7\times$ slower than \sysname, showing that L2 cache alone cannot fully eliminate redundant KV loads. In addition, DeFT and Cascade Inference both use a naive packing scheme for shared prefixes. While DeFT's load-balancing reduces SM execution bubbles from long-tail CTAs, neither method effectively reduces global memory accesses, which dominate decode attention latency. Therefore, \sysname achieves $76.6\%$ and $41.2\%$ lower attention latency on average than DeFT and Cascade Inference, respectively.

%% file: content/5_3_eval_e2e.tex
\subsection{End-to-End Comparison}
\label{sec:eval_e2e}

\parab{Overall trends.} We compare \sysname with three baselines across two models (Llama-3-8B and Qwen3-8B) and two real-world traces (conversation and toolagent) as in \secref{sec:eval_setup}. \figref{fig:eval_e2e_overall} shows the trend of mean TTFT, mean TPOT, and P99 TPOT as we vary the request rate. Specifically, as in subfigures \num{b1} to \num{b4}, \sysname reduces mean TPOT at the same request rate by 17.2--68.1\% over RelayAttention++, 17.0--89.5\% over FlashAttention, and 32.2--93.1\% over FlashInfer. Furthermore, as shown by the first and the third rows of \figref{fig:eval_e2e_overall}, the TPOT reduction allows \sysname to finish incoming requests faster at the same request rate. This yields 9.3--98.6\%, 10.1--99.6\%, and 22.5--99.8\% lower TTFT than the three baselines.

\parab{Scaling with request rate.} The performance gap between \sysname and the baselines first widens and then slightly contracts as the request rate grows. This is because: (1) Higher request rate forms larger decode batches under continuous batching, which increases the redundant global memory accesses and exposes more opportunity for \sysname; (2) When the batch size further increases, the number of CTAs and the overall attention runtime grow, so that the execution bubble becomes smaller; the performance improvement of multi-stream forward and long-KV split therefore shrinks. 

\parab{Why baselines fail.} \sysname achieves significant gains for its prefix-aware pack scheduler, multi-tile kernel, and multi-stream forward. In contrast, RelayAttention++ supports only a single-level system prefix and delegates the forward kernel to FlashAttention, so its curves largely track FlashAttention, which can not mitigate redundant KV loads at all. Model configuration also matters. Llama-3-8B’s 8K context length limit (vs.\ 32K for Qwen3-8B) requires us to filter ultra-long requests, which reduces the requests with long non-prefix context and leads to lower absolute latency for both RelayAttention++ and \sysname on Llama-3-8B compared with that of Qwen3-8B. Besides, FlashInfer improves SM utilization via long-CTA-splitting load balance, which helps at low request rate while adding scheduling overhead that grows with request rate, resulting in the highest TPOT.


\subsection{Distributed and MoE Extensions}
\label{sec:eval_dist_moe}

\begin{figure}[t]
    \centering
    \includegraphics[width=\linewidth]{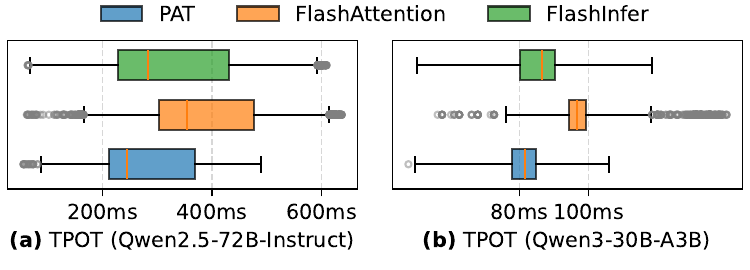}
    \caption[]{End-to-end performance of \sysname and baselines under TP/PP and MoE architectures.}
    \label{fig:eval_dist_moe}
    \Description{}
\end{figure}

Recent production LLM deployments increasingly rely on multi-GPU inference as model weights exceed a single GPU’s memory capacity, and many models adopt Mixture-of-Experts (MoE) architectures to increase capacity at moderate cost. \sysname seamlessly supports tensor parallelism (TP), pipeline parallelism (PP), and MoE: TP and PP partition attention heads or transformer blocks across GPUs, while each device retains full KV cache for its assigned heads or blocks; MoE architecture replaces the FFN layer with multiple experts but leaves attention execution unchanged. Therefore, \sysname's pack scheduler and multi-tile kernel run unchanged.

To validate these properties, we evaluate \sysname under the toolagent trace using Qwen2.5-72B-Instruct with TP=2 and PP=2 on four A100 GPUs, and Qwen3-30B-A3B on one A100 GPU. As shown in \figref{fig:eval_dist_moe}, compared to three baselines, \sysname reduces average TPOT by $14.3-26.7\%$ on Qwen2.5-72B-Instruct and by $5.53-16.9\%$ on Qwen3-30B-A3B. These results suggest that prefix-aware packing and the multi-tile kernel remain effective under common distributed and MoE deployment configurations.

%% file: content/5_4_eval_ablation.tex
\subsection{Ablation Study}
\label{sec:eval_ablation}

\begin{figure}[t]
    \centering
    \subfloat[attention latency]{
        \includegraphics[width=0.263\textwidth]{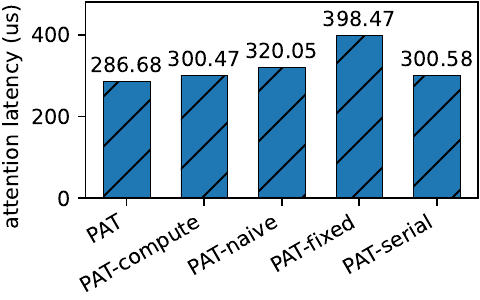}
        \label{fig:ablation_overall_latency}
    } 
    \subfloat[memory read/write]{
        \includegraphics[width=0.1972\textwidth]{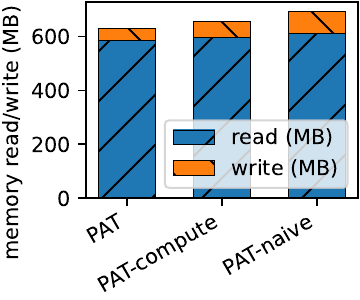}
        \label{fig:ablation_overall_memory}
    }
    \caption[]{Results of ablation baselines. (a) Average attention latency. (b) Global memory read and write size.}
    \label{fig:ablation_overall}
    \Description{}
\end{figure}

\begin{figure}[t]
    \centering
    \subfloat[multi-stream]{
        \includegraphics[width=0.23\textwidth]{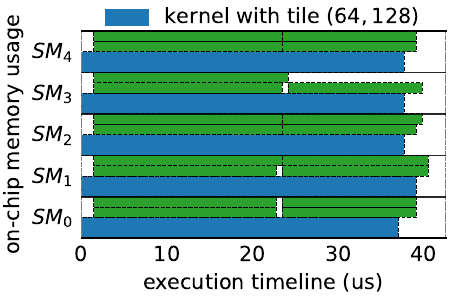}
        \label{fig:ablation_CTA_pipeline_multi}
    } 
    \subfloat[serial execution]{
        \includegraphics[width=0.23\textwidth]{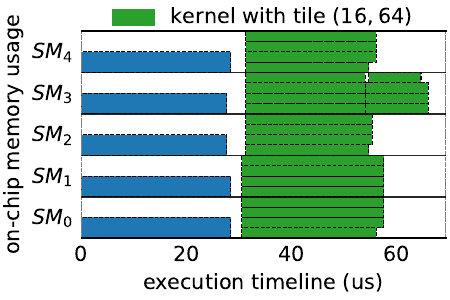}
        \label{fig:ablation_CTA_pipeline_serial}
    }
    \caption[]{CTA execution pipeline on $SM_{0}-SM_{5}$ with two tile size configurations (collected by PTX~\cite{nvidia2023ptxisa}), where white space represents execution bubbles. (a) Multi-stream execution pipeline. (b) Serial execution pipeline.}
    \label{fig:ablation_CTA_pipeline}
    \Description{}
\end{figure}

To evaluate the contribution of each design in \sysname, we build the following ablation baselines: 
(1) \sysname-compute, which adopts the cost model from FastTree~\cite{pan2025fasttree} for the packing scheduler; 
(2) \sysname-naive, which simply packs each node in the tree-structured block table into a CTA;
(3) \sysname-fixed, which disables the multi-tile kernel of \sysname and instead uses the fixed tile configuration $(64, 128)$ as in FlashAttention~\cite{shah2024flashattention3};
(4) \sysname-serial, which disables the multi-stream forward mechanism of \sysname and adopts the serial multi-kernel execution similar to FastTree~\cite{pan2025fasttree}.
We use the same synthetic traces as in \secref{sec:eval_kernel_perf} and adopt the attention head configuration of Llama-3-8B to compare \sysname with ablation baselines.

\parab{Effectiveness of the pack scheduler.}
As shown in \figref{fig:ablation_overall_latency}, the average attention latency of \sysname-compute and \sysname-naive is higher than \sysname by 4.6\% and 10.4\%, respectively. \sysname-compute adopts a compute-oriented cost model that selects the packing scheme with the minimum computation, which contradicts the memory-bound nature of decode attention. Meanwhile, \sysname-naive only considers the benefit of packing but ignores the additional intermediate reads and writes, leading to high overhead. As shown in \figref{fig:ablation_overall_memory}, their average memory read/write is higher than \sysname by 10.9\% and 16.7\%, respectively. This confirms the rationality of \sysname’s memory-oriented cost model and heuristic pack scheduler.

\parab{Importance of multi-tile kernel.}
As shown in \figref{fig:ablation_overall_latency}, enforcing a fixed tile size (\sysname-fixed) increases attention latency by 39\% compared to \sysname. In the ablation workload, CTAs exhibit query sizes from 1 to 64 and KV lengths from 32 to 4096, which makes one-size-fits-all kernels highly inefficient. Prior kernels~\cite{shah2024flashattention3, ye2025flashinfer, zhu2024relayattention, yao2024deft} adopt such fixed designs, leading to padding overhead and execution bubbles. In contrast, \sysname's multi-tile kernel enables efficient adaptation to diverse CTA configurations.

\parab{Effectiveness of multi-stream forward.}
As shown in \figref{fig:ablation_overall_latency}, the average attention latency of \sysname-serial is 4.8\% higher than \sysname. This is because serial execution aggravates execution bubbles. For example, in a two-level prefix decode batch (\figref{fig:eval_kernel_overall} \num{6}), \figref{fig:ablation_CTA_pipeline_multi} and \figref{fig:ablation_CTA_pipeline_serial} show the CTA execution pipelines of \sysname and \sysname-serial, where \sysname-serial suffers from substantial memory waste and execution bubbles. At runtime, \sysname's multi-tile kernel selects a tile configuration for each CTA from the equivalent configuration set. In our A100 experiments, each batch typically uses 1–5 of the 11 available configurations (\figref{fig:design_pack.kernel}). \sysname's multi-stream forward then launches a separate CUDA stream for CTAs with each active tile configuration, allowing these streams to run in parallel under the GPU scheduler. With multiple configurations, some launch bubbles remain, as indicated by the blank region on the left of \figref{fig:ablation_CTA_pipeline_multi}. Nevertheless, parallel streams substantially reduce execution bubbles compared to serial execution, thereby lowering attention latency.



%% file: content/5_5_eval_overhead.tex
\subsection{Overhead Analysis}
\label{sec:eval_overhead}

\begin{figure}[t]
    \centering
    \includegraphics[width=\linewidth]{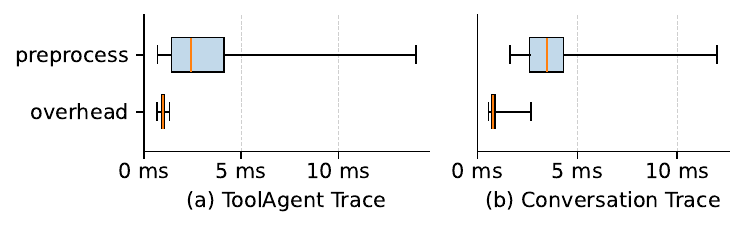}
    \caption{Overhead of pack scheduler and pre-attention task latency of serving system.}
    \label{fig:eval_overhead}
    \Description{}
\end{figure}

The primary overhead introduced by \sysname comes from the pack scheduler, which packs decode batches into CTAs at runtime. \sysname's lazy update mechanism (\secref{sec:design_pack.pack_scheduler}) is designed to mitigate the overhead. It reduces the scheduler's triggering frequency without affecting correctness and overlaps its execution with the serving system's pre-attention tasks (\eg metadata preparation, QKV projection). To validate, we measure both the scheduling latency and the pre-attention task latency under toolagent and conversation traces at request rates of 5 and 8 req/s. As shown in \figref{fig:eval_overhead}, the average scheduling latency is lower than the pre-attention task latency by 42.3\% and 49.6\%, respectively. Consistently, the P99 TPOT in \figref{fig:eval_e2e_overall} \num{c1}-\num{c4} shows 19.4–93.4\% reduction compared with the baselines. Therefore, when running on an asynchronous CPU thread, the pack scheduler will \textit{not} introduce additional end-to-end latency, demonstrating the effectiveness of the lazy update mechanism in \sysname.

%% file: content/6_related_conclusion.tex
\section{Discussion}
\label{sec:discussion}

\parab{Prospects and Limitations.} \sysname leverages cross-request prefix sharing to improve bandwidth utilization and cut redundant global memory accesses in memory-bound decode attention. Its effectiveness depends on three factors: (1) \textit{hardware compute-to-bandwidth ratio.} As GPUs become more compute-dominant (\eg NVIDIA V100 to B200: 139 to 312 FLOP/Byte), the gap between compute and memory widens, so memory-focused designs like \sysname become increasingly valuable; (2) \textit{model architecture.} \sysname yields large gains for common architectures that retain a KV cache (MHA, GQA), but benefits may shrink for architectures that compress or remove KV state (MLA~\cite{liu2024deepseek}, linear attention~\cite{shen2021efficient}, MLKV~\cite{zuhri2024mlkv}); and (3) \textit{prefix-sharing ratio in the batch.} High concurrency with cross-request shared prefixes amplifies \sysname’s advantage, whereas small batches or workloads without shared prefixes limit the improvement (see \figref{fig:eval_kernel_overall} \num{1}, \num{19}–\num{20}).

\parab{Gap to Optimal.} \sysname reduces memory waste and execution bubbles, guided by hardware and pipeline analysis. However, GPU scheduling is uncontrollable, leaving residual bubbles (\figref{fig:ablation_CTA_pipeline_multi}) and a gap from the theoretical optimum. Yet \sysname consistently outperforms state-of-the-art baselines.

\section{Related Work}
\label{sec:related}

\parab{KV Cache Related Optimization.}
Prior work reduced KV cache memory and fragmentation: FasterTransformer~\cite{nvidia2025fastertransformer} provided an early static-batching implementation, Orca~\cite{yu2022orca} improved utilization with continuous batching but relied on pre-allocated caches, which leads to memory fragmentation. vLLM introduced a paged KV cache to cut waste by virtual memory management~\cite{kwon2023pagedattention}. SGLang's prefix-reuse lets shared prefix KV blocks be reused across requests and is widely used in production~\cite{zheng2024sglang}. Recent systems~\cite{qin2024mooncake, lee2024infinigen, hu2025tightllm} further extend KV capacity by offloading KV blocks between GPU and CPU or NVMe storage, which reduces on-GPU memory pressure at the cost of additional data movement. These approaches shrink KV memory costs but do not accelerate decode attention, which dominates latency as context and output lengths grow. \sysname is orthogonal to these approaches because its pack scheduler only relies on logical KV block IDs, and the serving system transfers the required KV blocks to GPU memory before the attention kernel executes.

\parab{Attention Kernel Optimization.}
Fused, on-chip attention kernel implementations like FlashAttention and FlashInfer reduce global memory traffic by combining attention steps into a single kernel~\cite{dao2022flashattention, shah2024flashattention3, ye2025flashinfer}, but their query-centric (one-query-per-CTA) design cannot exploit workload-level prefix sharing. Subsequent works~\cite{zhu2024relayattention, ye2024chunkattention, lin2024parrot, zheng2024batchllm} pack the single shared system prompt to reduce redundant memory accesses, but can not generalize to workloads with multi-level prefixes. Several recent works~\cite{zihao2024cascade, yao2024deft, juravsky2024hydragen} extend to multi-level prefixes but rely on simple packing strategies that overlook trade-offs between overhead and savings. \cite{pan2025fasttree} also addresses multi-level prefix optimization, but its compute-oriented packing cost model mismatches the memory-bound nature of decode attention, leading to suboptimal results. Furthermore, all these works use one-size-fits-all kernel design, leading to resource inefficiency. In contrast, \sysname uses a memory-centric packing strategy with multi-tile kernels and multi-stream execution to reduce redundant memory accesses and improve resource utilization for dynamic workloads.

\parab{GPU Scheduling.}
GPU scheduling strategies further improve kernel efficiency by balancing compute and memory demands: \cite{kamath2025pod} fuses prefill and decode via virtual CTAs to expose resource complementarity, and \cite{zhu2025nanoflow} groups SMs so different SMs specialize in specific tasks, raising utilization. These techniques are largely orthogonal to \sysname and could be combined to further reduce execution inefficiencies. Work like \cite{song2024tackling} shows that CTA-style scheduling on alternative accelerators (\eg NPUs) can help eliminate the remaining resource bubbles in \sysname’s execution (\secref{sec:discussion}). Here, we focus on GPU decode-attention optimizations and leave heterogeneous-scheduler integration to future work.

\section{Conclusion}
\label{sec:conclusion}

In this work, we present \sysname, a prefix-aware attention kernel implementation for LLM decoding. \sysname reduces redundant global memory accesses and improves bandwidth utilization through a pack–forward–merge paradigm. The design incorporates a heuristic pack scheduler, a resource-efficient multi-tile kernel, and forward-stage optimizations that mitigate execution bubbles. We implement \sysname as an off-the-shelf plugin to vLLM and demonstrate that it reduces the attention latency by 53.5\% on average under synthetic workloads and reduces the TPOT by up to 93.1\% under real-world workloads compared with existing works.

\section{Acknowledgments}

We appreciate the insightful feedback from the anonymous reviewers and our shepherd, Seonjin Na.
This work is supported by the National Natural Science Foundation of China under grants No. 62572341 and No. 62202328.

%% file: content/_appendix.tex
\clearpage
\appendix

\twocolumn[{
\begin{center}
    \includegraphics[width=\textwidth]{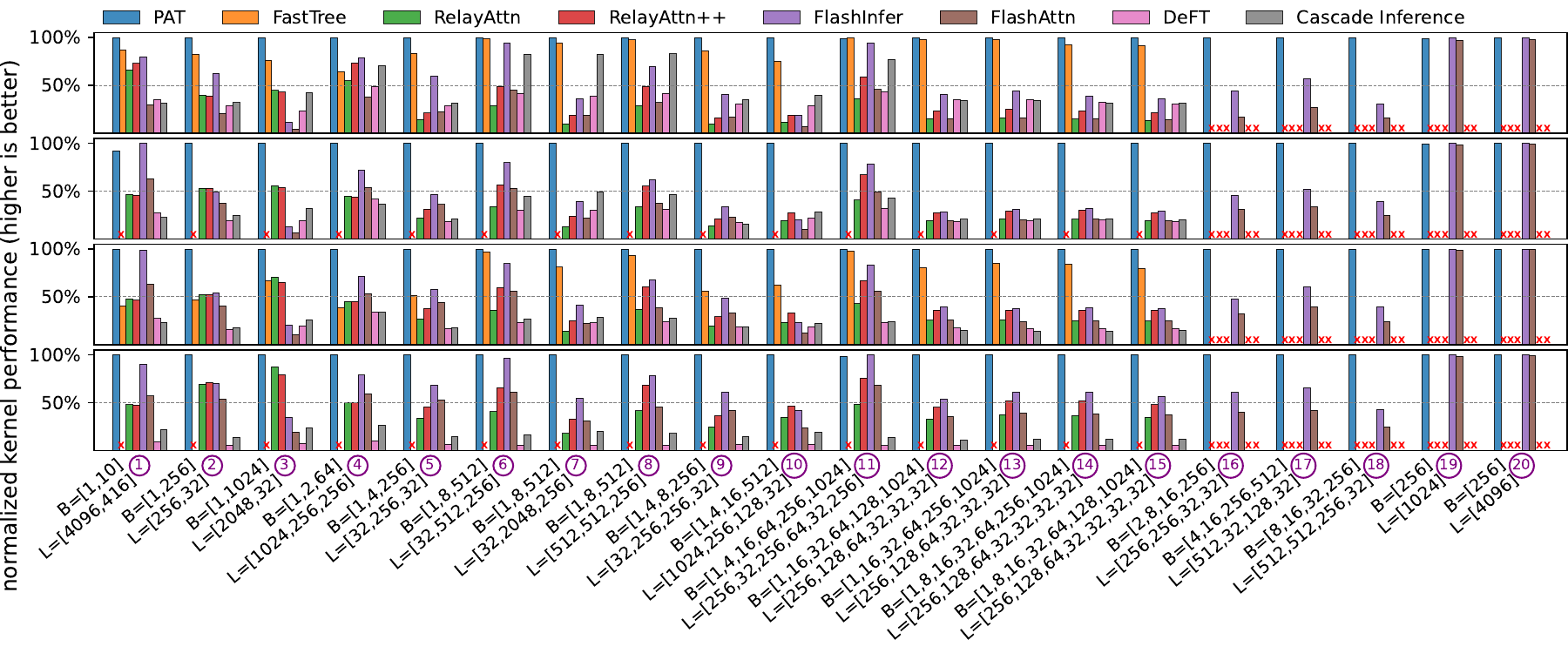}
    \captionof{figure}{Normalized kernel performance (higher is better) of \sysname and the baselines for the attention computation across various decode batch configurations on NVIDIA H100 GPU. The four panels from top to bottom show head configurations ($num\_attention\_heads/num\_key\_value\_heads$) of 32/32, 16/8, 32/8, and 64/8. Missing bars arise because RelayAttention lacks support for multi-level or multiple first-level prefixes, and FastTree does not support the 16/8 and 64/8 head settings.}
    \label{fig:appendix_kernel_overall}
\end{center}
}]


\section{Kernel Performance on H100 GPU}
\label{sec:appendix_kernel_perf}

\sysname's multi-tile kernel adapts to different GPU architectures by re-deriving equivalent tile-size configurations from the three constraints in \secref{sec:design_pack.multi_tile}. This allows the kernel to match different memory hierarchies and bandwidths. To validate this, we compare \sysname with the baselines in \secref{sec:eval_setup} on an NVIDIA H100 GPU, using configurations in \figref{fig:eval_kernel_overall}. As in \figref{fig:appendix_kernel_overall}, \sysname achieves kernel performance of $1.3-6.9\times$ to baselines with shared prefixes (\num{1}-\num{18}), and achieves consistent performance without shared prefixes (\num{19} and \num{20}). These results further confirm the robustness of \sysname's design.

\section{Artifact Appendix}
\label{sec:appendix_artifact}

\subsection{Abstract}

We provide the source code of \sysname along with scripts to reproduce experimental results presented in \secref{sec:eval}. This appendix includes instructions for reproducing the kernel performance evaluation results in \figref{fig:eval_kernel_overall} and four representative end-to-end performance evaluation results in \figref{fig:eval_e2e_overall}. 

To expedite artifact evaluation, a pre-built Docker image is available that contains the fully configured environment, pre-compiled source code, and datasets as specified in \secref{sec:eval_setup}. The experiment require an x86-64 Linux host with at least 64GB RAM, 200GB of free disk space, and an NVIDIA A100 GPU (80GB). For convenience and consistent performance, we recommend using a Google Cloud a2-ultragpu-1g instance with the “Deep Learning VM with CUDA 12.4” system image.


\subsection{Artifact check-list (meta-information)}

{\small
\begin{itemize}
  \item {\bf Algorithm: } Prefix-aware attention kernel implementation.
  \item {\bf Compilation: } Pre-compiled with in a Docker image.
  \item {\bf Model: } Llama3-8B and Qwen3-8B.
  \item {\bf Run-time environment: } Docker, NVIDIA Container Toolkit, and CUDA driver $\ge$ 550.
  \item {\bf Hardware: } 1$\times$ NVIDIA A100-80GB GPU.
  \item {\bf Metrics: } kernel latency, mean TTFT, mean TPOT, P99 TPOT.
  \item {\bf Output: } Figure and console.
  \item {\bf Experiments: } Kernel performance (\figref{fig:eval_kernel_overall}); End-to-end performance (\figref{fig:eval_e2e_overall}).
  \item {\bf How much disk space required (approximately)?: } 200GB.
  \item {\bf How much time is needed to prepare workflow (approximately)?: } 30 minutes.
  \item {\bf How much time is needed to complete experiments (approximately)?: } $\approx$2 hours for kernel performance experiment (\figref{fig:eval_kernel_overall}); $\approx$10 hours for the selected end-to-end performance experiments (\figref{fig:eval_e2e_overall}).
  \item {\bf Publicly available?: } Yes.
  \item {\bf Code licenses (if publicly available)?: } MIT.
  \item {\bf Archived (provide DOI)?: } \href{https://doi.org/10.5281/zenodo.18217189}{10.5281/zenodo.18217189}.
\end{itemize}
}

\subsection{Description}

\subsubsection{How to access} 
\begin{itemize}
    \item GitHub: \href{https://github.com/flashserve/PAT}{https://github.com/flashserve/PAT}
    \item Zenodo: \href{https://doi.org/10.5281/zenodo.18217189}{10.5281/zenodo.18217189}
    \item Pre-complied Docker image: \href{https://hub.docker.com/layers/flashserve/pat/ae}{flashserve/pat:ae}
\end{itemize}

\subsubsection{Hardware dependencies} Requires an x86-64 Linux host with at least 64GB of RAM, 200GB of free disk space, and an NVIDIA A100 GPU (80GB). For convenience and consistent performance, we recommend using a Google Cloud a2-ultragpu-1g instance with the "Deep Learning VM with CUDA 12.4" system image.

\subsubsection{Software dependencies} Docker, NVIDIA Container Toolkit, and NVIDIA driver $\ge$550 are required. Other software has been installed within the provided Docker image.

\subsubsection{Data sets} The data sets used in experiments are listed in \secref{sec:eval_setup} and are pre-downloaded in the Docker image.

\subsubsection{Models} The model used in these experiments are listed in \secref{sec:eval_setup}, including Qwen3-8B and Llama-3-8B.

\subsection{Installation}

\begin{enumerate} 

\item Clone the GitHub repository.
\begin{lstlisting}[language=bash]
$ git clone 
  https://github.com/flashserve/PAT.git
\end{lstlisting}

\item Pull the pre-built Docker image.
\begin{lstlisting}[language=bash]
$ # about 50GB, including model weights
$ docker pull flashserve/pat:ae
\end{lstlisting}

\item Start a Docker container with GPU access and mount the repository.
\begin{lstlisting}[language=bash]
$ docker run -it --gpus all --shm-size=64g \
    -v ${PWD}/PAT:/workspace/PAT \
    -w /workspace/PAT \
    flashserve/pat:ae /bin/bash
\end{lstlisting}

\end{enumerate} 

\subsection{Experiment workflow}

\begin{enumerate} 

\item Run the kernel performance experiments. This experiment takes about 1.5 hours to complete.
\begin{lstlisting}[language=bash]
$ cd /workspace/PAT/benchmark
$ bash ./run_kernel_bench.sh
\end{lstlisting}

\item Run the end-to-end serving performance experiments. Note that completing all experiments requires over 60 GPU-hours, so we provide two scripts for convenience: (1) \texttt{run\_e2e\_bench\_part.sh}: runs a subset of experiments (QPS=7\&9, all workloads, all baselines) for quick verification; (2) \texttt{run\_e2e\_bench\_full.sh}: runs all experiments in \figref{fig:eval_e2e_overall}.
\begin{lstlisting}[language=bash]
$ cd /workspace/PAT/benchmark
$ # Quick verification (8-10 GPU-hours)
$ bash ./run_e2e_bench_part.sh
$ # Full experiments (over 60 GPU-hours)
$ # bash ./run_e2e_bench_full.sh
\end{lstlisting}

\end{enumerate} 

\subsection{Evaluation and expected results}

Generate plots. 

\begin{lstlisting}[language=bash]
$ cd /workspace/PAT/plot
$ python eval_kernel_perf.py --log-file \
  ../benchmark/kernel_perf.json
$ python eval_e2e_from_jsonl.py --log-file \
  ../benchmark/e2e_perf.jsonl
\end{lstlisting}
